# Securely extending and running low-code applications with C#

submitted by

Lennart Brüggemann, B.Sc.

lennart.brueggemann@studium.fernuni-hagen.de

Matriculation number 9767045

# Eidesstattliche Erklärung

Ich erkläre, dass ich die Masterarbeit selbstständig und ohne unzulässige Inanspruchnahme Dritter verfasst habe. Ich habe dabei nur die angegebenen Quellen und Hilfsmittel verwendet und die aus diesen wörtlich, inhaltlich oder sinngemäß entnommenen Stellen als solche den wissenschaftlichen Anforderungen entsprechend kenntlich gemacht. Die Versicherung selbstständiger Arbeit gilt auch für Zeichnungen, Skizzen oder graphische Darstellungen. Die Arbeit wurde bisher in gleicher oder ähnlicher Form weder derselben noch einer anderen Prüfungsbehörde vorgelegt und auch noch nicht veröffentlicht. Mit der Abgabe der elektronischen Fassung der endgültigen Version der Arbeit nehme ich zur Kenntnis, dass diese mit Hilfe eines Plagiatserkennungsdienstes auf enthaltene Plagiate überprüft und ausschließlich für Prüfungszwecke gespeichert wird.

# Statement in lieu of an oath

I hereby confirm that I have written this thesis on my own and without unauthorized help of third parties. I have used only the indicated sources and tools and have marked all matter taken from them, verbatim, content-related or paraphrased, in accordance to scientific requirements. The declaration of independent work also applies to drawings, sketches or graphical representations. This work or parts thereof have not been presented to the same or any other examination authority and has not yet been published. By submitting the electronic form of the final version of the thesis I acknowledge that it will be checked for plagiarism with the help of a plagiarism detection service and stored exclusively for examination purposes.

______________________________  ______________________________
Place/Date                              Signature

*For Ruth and Georg*

# Abstract

Low-code development platforms provide an accessible infrastructure for the creation of software by domain experts, also called 'citizen developers', without the need for formal programming education. Development is facilitated through graphical user interfaces, although traditional programming can still be used to extend low-code applications, for example when external services or complex business logic needs to be implemented that cannot be realized with the features available on a platform.

Since citizen developers are usually not specifically trained in software development, they require additional support when writing code, particularly with regard to security and advanced techniques like debugging or versioning. In this thesis, several options to assist developers of low-code applications are investigated and implemented. A framework to quickly build code editor extensions is developed, and an approach to leverage the Roslyn compiler platform to implement custom static code analysis rules for low-code development platforms using the .NET platform is demonstrated. Furthermore, a sample application showing how Roslyn can be used to build a simple, integrated debugging tool, as well as an abstraction of the version control system Git for easier usage by citizen developers, is implemented.

Security is a critical aspect when low-code applications are deployed. To provide an overview over possible options to ensure the secure and isolated execution of low-code applications, a threat model is developed and used as the basis for a comparison between OS-level virtualization, sandboxing, and runtime code security implementations.

# Contents

# List of Acronyms

| | |
|---|---|
| CVE | Common Vulnerabilities and Exposures |
| VSIX | Visual Studio integration extension |
| WPF | Windows Presentation Foundation |
| WYSIWYG | What You See Is What You Get |
| XAML | Extensible Application Markup Language |

# List of Abbreviations

| | |
|---|---|
| ACL | access control list |
| API | application programming interface |
| CD | citizen developer |
| CMSA | code management sample application |
| DSL | domain specific language |
| GUI | graphical user interface |
| GUID | globally unique identifier |
| IDE | integrated development environment |
| IL | integrity level |
| IPC | inter-process communication |
| LCA | low-code application |
| LCDP | low-code development platform |
| MAC | mandatory access control |
| MDM | mobile device management |
| MIC | mandatory integrity control |
| OS | operating system |
| SDK | software development kit |
| SID | secure identifier |
| UI | user interface |
| VCS | version control system |
| VM | virtual machine |

# Glossary

Assembly    A library or application that contains types and resources for use in .NET, typically a DLL or EXE file.

blob    An unstructured, binary representation of a data object, such as a picture or audio file. Today commonly referred to as *binary large object* or *basic large object* [145].

NuGet    Package manager for .NET, developed by Microsoft. Packages frequently contain internal or external libraries that can be installed into a .NET project.

Roslyn    Compiler platform and SDK for .NET, developed by Microsoft. Provides APIs to modify, compile and interact with source code.

# Chapter 1

# Introduction

## 1.1 Background

Low-code development platforms (LCDPs) aim to decrease development and testing cycles of software, and reduce the need for software engineers to create software using traditional programming workflows [135]. Demand outweighs the availability of professional developers in both Germany [155] and the United States [150], thus companies try to fill their positions with domain experts without formal programming experience to develop their software [47]. These professionals are also called citizen developers (CDs), and, while having none or very little experience in software development, they are skilled and experienced in their particular domain [158].

LCDPs enable such experts to develop applications in their area of knowledge, the only additional training necessary is for the platform used itself. These tools usually use some variation of a *What You See Is What You Get* (WYSIWYG) interface to build the user interface (UI) for an application [60]. For example, by dragging and dropping buttons, text boxes, tables, or images onto a surface that will resemble the front-end later on [135, p. 15], and then setting properties like colours or text using simple editors. This contrasts with manual development of front-end interfaces, which involve programming a UI either fully in code, or combining it with a description language such as XAML (Extensible Application Markup Language). WYSIWYG-editors exist as part of integrated development environments (IDEs), such as Visual Studio or Eclipse, too, but still directly generate code for the developer to customize and integrate with the rest of the code base. In comparison, a low-code developer has little to no direct interaction with source code. However, what happens if the low-code platform does not provide enough possibilities to cover a domain-specific use case, such as the inability to express a certain logic, or limited options for UI-design?

LCDPs still allow a varying degree of customization with additional manually written code (*No-code* platforms exist as well [60]). For example, the Mendix platform supports extension of their applications via so-called *Java actions* [83]. Microsoft's Power Apps platform allows for additional *code components* [22] to be implemented. How custom code is used by the low-code platform differs between each of them. Some platforms generate native applications for the target system (Windows applications, Android or iOS apps), for example Mendix or Scopeland [141]. The user-written code is part of the native application in this case. Others run inside the low-code environment itself, such as the aforementioned PowerApps [29]. During this thesis, the former type of resulting application will be called *generated application*, the latter *hosted application*. The different run-time environments present unique challenges in terms of security, debugging and auditing the resulting code.

## 1.2 Motivation

Traditionally developed software has numerous tools, techniques, and patterns available to tackle issues concerning the mentioned challenges. Debuggers exist standalone and as part of IDEs, security policing is possible through static code analysis during development time, and auditing can be facilitated by rigorous internal or external examination of the written source code. However, in low-code environments, the resulting source code is not as readily accessible, since it is only generated when needed. While the code generators of the low-code platform itself can be accessed and evaluated by normal means (since they were usually developed conventionally), the supplemental code written by citizen developers will become either part of the generated application or, in case of hosted applications, be executed during run-time. Adding or running code in an unchecked manner can present both security and maintenance issues.

The Open Web Application Security Project (OWASP)[1], a non-profit organization known for publishing the OWASP Top 10 [126] (a list of the ten most common security risks for web applications), issued a first draft of the Top 10 for low- and no-code applications in 2021 [8]. Especially the risks '*LCNC-SEC-06: Injection Handling Failures*' and '*LCNC-SEC-07: Vulnerable and Untrusted Components*' highlight the possible security vulnerabilities that can be incurred by adding additional code to a low-code application.

Supporting citizen developers in writing code, if they want or have to, is another aspect to consider. Instead of only offering the built-in application programming interface (API) provided by the standard library a particular low-code platform uses (like the base class libraries of .NET or Java), additional APIs can be provided. Such custom APIs may allow simplified access to the environment the code will run in (like the currently developed application or the platform itself), or a façade against the standard library (e.g., easier access to image manipulation). Providing more approachable access to common software engineering tools like version control systems such as Git or Subversion also helps in defect finding and auditing, another security hazard highlighted in the OWASP Top 10 as '*LCNC-SEC-10: Security Logging and Monitoring Failures*' [8].

---

[1] https://owasp.org

This thesis will investigate the challenges and problems of user-written code in low-code platforms. Specifically, code written in the C# programming language and utilizing the .NET development framework will be the focus of the programming aspect. Possible solutions in terms of integrating user-written C# code into the low-code development environment, and its secure execution, debugging and revision will be examined. The platform SCOPELAND, further introduced in section 2.4, will be used as an example low-code platform. However, the results are not aimed towards a specific product and are discussed in a generalized form.

## 1.3 Related work

Most of the work in the area is relatively recent, according to multiple sources [46, 75, 158] the term *low-code* was coined in 2014 in a report by the market research company Forrester Research [130]. There has been work on the categorization and comparability of low-code development platforms, for example, Sahay et al. present a taxonomy for LCDPs and classified eight of them [134]. Their categorization is primarily based on concrete features that are required to build an application and the features offered by the examined LCDPs.

Frank et al. [46] assessed ten platforms and examined them by their common characteristics and features. They use a more theoretical approach based on a model that was defined before the examination of concrete platforms to prevent bias towards a specific vendor [46, p. 5]. In comparison to the other mentioned work, they also evaluate how the vendors market and place their products. As part of their assessment of the extensibility with additionally written code for each platform, they considered the use of source code editors as well (if available), and highlighted problems with their usage and how users were supported in their use. As a key finding, they state that '*In general, the more code editing is enabled by an LCDP, the more demanding is its use*' [46, p. 158], which lends itself to additional research, namely how to make code editing in LCDPs *less* demanding.

The understanding and challenges of LCDPs from the perspective of developers was explored by Luo et al. [75] by analysing posts in online communities. Their approach leans on the extraction of data from posts on StackOverflow[2] and from certain communities on Reddit[3], which they use to answer several research questions to classify LCDPs based on the experience of practitioners.

Testing of applications developed in LCDPs was analysed by Khorram et al. [68], including an evaluation of the support for testing in five low-code platforms and the role of the citizen developer in the creation of tests for their developed applications. Their article also highlights the need to support citizen developers not only in developing an application, but also in further aspects of software development — in this case testing. For the OutSystems platform, Jacinto et al. [64] present an approach for mock-based unit testing. Although not described in detail, their approach of changing the way the OutSystems' code generator works, adding control flow statements to call the mock code during runtime, is a form of *interception* also used in aspect-oriented programming. This is interesting because it shows that patching generated code can be an option in LCDPs to alter the resulting application at a low level during runtime.

---

[2] A QA-platform for professional and enthusiast software developers. Strict moderation, a voting-system and wiki-style editing of other users' questions and answers aim to create a database of high-quality canonical answers to problems — `https://stackoverflow.com`

[3] A social media platform for discussions of almost every topic. Discussions are thematically ordered in *subreddits* (a kind of forum), which every user can start and manage — `https://reddit.com`

Outside of publications by the respective vendors, there is little independent research on the security of low-code platforms or their applications. Oltrogge et al. [122] analyse the security of Android apps generated by *online application generators*, by auditing the generated boilerplate code and identifying possible attack vectors and checking them against Google's security best practices for developing apps.

## 1.4 Research Questions

Assistance for citizen developers in writing code for LCAs varies between platforms, and often aims at more professional developers [46, p. 159]. The first question is thus derived from the issue, how to make writing code more approachable for citizen developers:

> *RQ I: How can citizen developers in low-code environments that offer C#-based extension of their applications be supported?*

While citizen developers are experienced in their particular domain, they may only have basic training in programming are not expected to have a thorough understanding of software architectures, development patterns and best practices. Supporting them in writing better extensions may not be enough, as, no matter the profession, experience or education, code will always contain minor and major bugs. Since security is one of the perceived concerns in LCDPs [158], a platform must also actively address potential security issues that can arise from custom code. This leads to the second research question:

> *RQ II: How to guard against malicious or unintended harmful C# code that is written to extend low-code applications?*

However, security problems don't exclusively stem from additional code written for a LCA. Vulnerabilities can also exist in external libraries used in the LCA or come by misconfiguration of data transport or storage mechanisms. The last research question deals with the collaborative usage of LCDPs, and how to minimize the risk of interference when multiple clients share a server for developing LCAs:

*RQ III: How can the execution of low-code applications be isolated in a way, that additionally written C# code cannot adversely affect foreign resources on a system?*

## 1.5 Results and contributions

This thesis explores improvements to the usability of code editors for citizen developers in low-code applications. For this purpose, several libraries and example applications were developed by the author:

- A framework that can be used to add custom text decorations to the AvalonEdit code editor is presented in section 3.3. This library aims to enable the implementation of as many visual augmentations, as defined and categorized by Sulír et al. [146], as possible.

- The framework also provides functions to implement the display of context-based *advice*, that supports users in the correct usage of APIs. These advice messages are based on the work of Gorski et al. [54] and detailed in section 3.4.2.

- An application that shows the benefits of wrapping existing, difficult to use APIs, derived by the work of Jugel et al. [67], was developed as part of section 3.4.1.

- An example application that demonstrates the use of an AppContainer, an implementation of process-level isolation using mandatory integrity control (MIC) on Windows, was developed as part of section 4.5.3.

- A code management sample application (CMSA) demonstrating on-demand compilation of C# scripts during runtime using the Roslyn compiler platform [93] and showing basic debugging possibilities using *printf debugging* [11] is introduced in section 5.3.1.

- The CMSA also shows how the version control system Git can be abstracted to make it more accessible for citizen developers, providing basic versioning of code, including a history and retrieval of older versions. The implementation is detailed in section 5.3.2.

## 1.6 Overview

Chapter 2 provides an overview of low-code technology and introduces the terminology used to distinguish between the types of LCAs and LCDPs. The chapter also explains how the extension of LCAs works in practice and presents the SCOPELAND platform, which is used as an example LCDP throughout the thesis.

Support mechanisms for CDs, based on static code analysis and extension of code editors, are presented in chapter 3. It also describes the implementation of a framework that enables the development of context-based visual augmentations for code editors in LCDPs, as well as the integration of custom static code analysis rules and automated code fixes.

The security aspect of LCDPs is covered in chapter 4. Based on a threat model, two approaches using operating system (OS)-level virtualization to run applications in an isolated environment are examined and evaluated. Furthermore, currently available mechanisms to secure individual processes with C# are discussed, and the implementation of a process-level container using Windows' built-in mandatory access control (MAC) is detailed.

Chapter 5 describes how two commonly used techniques in traditional software development, version control and debugging, can be simplified and integrated into existing LCDPs to make them more accessible for CDs. Based on a specifically developed sample application, an abstraction of the version control system Git and the integration of the Roslyn compiler toolkit to enable simple debugging features is demonstrated.

Chapter 6 summarizes the presented solutions in relation to the research questions posed previously. In addition, several ideas for further research and possible extensions to the applications implemented as part of this thesis are proposed.

# Chapter 2

# Low-code

Classifying the types of low-code platforms is a necessary step to contextualize problems and solutions, as not all of them can be applied universally to all low-code approaches. There have been efforts to categorize both low-code application (LCA) and their LCDPs, as described in section 1.3. Since this thesis doesn't seek to compare or analyse existing low-code platforms by features or specific vendors, a different classification based on technical characteristics is introduced. The first distinction made is between the environment the developed low-code application is running — either as part of a host platform or as a generated, native application on the system it is running on. The former approach can be further distinguished, by whether the host platform runs on a local computer or on a remote server (possibly in the cloud).

These characteristics are not mutually exclusive, for example, a LCDP may support both hosted and generated applications.

## 2.1 Types of low-code development platforms

Figure 2.1 shows the relationship between LCDPs and LCAs. LCDPs can be run either locally, as a regular application on the development system, or remotely on a web server or in the cloud. LCAs are built with LCDPs, but can run in different environments: hosted inside the LCDP itself, running in a separate application that serves as a *parent* for the application, or as a generated application, that runs natively on the target platform.

### 2.1.1 Local

Installing or deploying the LCDP locally on a computer also involves setting up the environment required for developing LCAs. Depending on the types of applications that will be built, this may require further components, such as drivers or software for debugging [125]. This can be a limiting factor for organizations, if installing third-party software is subject to security guidelines, such as the German IT-Grundschutz ('IT base security') compendium [21, Sec. APP.6.A4], and would require thorough analysis of any installed programs. Since the LCDP would have to be installed on all devices that will be used for development, additional administrative effort is required.

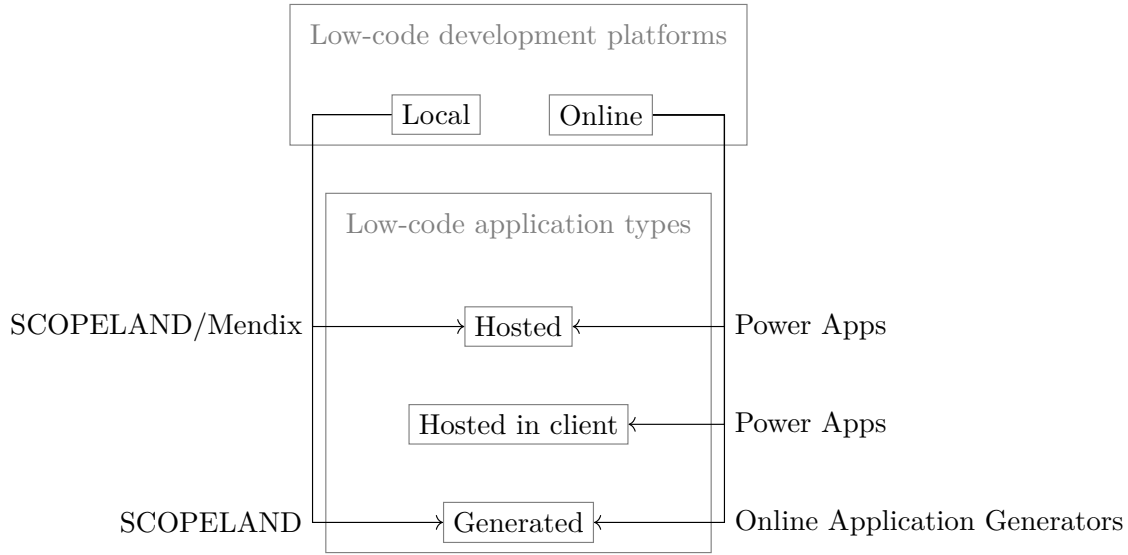

Figure 2.1: Examples for LCDPs with different LCA output types

Unlike web applications, locally run LCDPs (or more generally: desktop applications) have the advantage of having access to native platform APIs, such as Win32 (for Windows) or Cocoa (for macOS) and are not limited to only those APIs available to applications running in a browser [138]. This enables direct access to system resources and hardware, which are otherwise only available through specific APIs such as WebUSB [154].

Examples for local LCDPs are Mendix Studio Pro [81] or SCOPELAND Direct Desk, the latter being examined in section 2.4.

### 2.1.2  Online/Cloud

Running a LCDP online removes the need to deploy the development environment to the user's devices, as only a browser is required to run the platform. An implementation of cloud-based LCDPs is Microsoft's *Power Apps* platform. Figure 2.2 shows a page of the *Asset Checkout* sample app in its web-based designer. The same page, during runtime and populated with concrete data and running on Windows 11, is shown in figure 2.3.

These kinds of LCDPs are often provided through a Platform-as-a-Service (PaaS) model running in the cloud [134], although some vendors offer platforms running in a private cloud, such as Mendix [82]. While a web-based designer doesn't require specific deployment to the devices it will be used on, it presents a different set of issues not equally present in locally run applications: possible privacy and security problems.

Working with and storing data in the (public) cloud, especially when the data originates from within the European Union (EU) and is expected to flow to non-EU jurisdictions, can present legal challenges [157, p. 16] for organizations and may be undesirable for sensitive company or user data. Depending on the jurisdiction, including a cloud service into a company's infrastructure can also involve additional steps to satisfy security guidelines, such as those published by the German Federal Office for Information Security [143], and thus possibly negate advantages in time and effort gained by a system that does not need to run on premises.

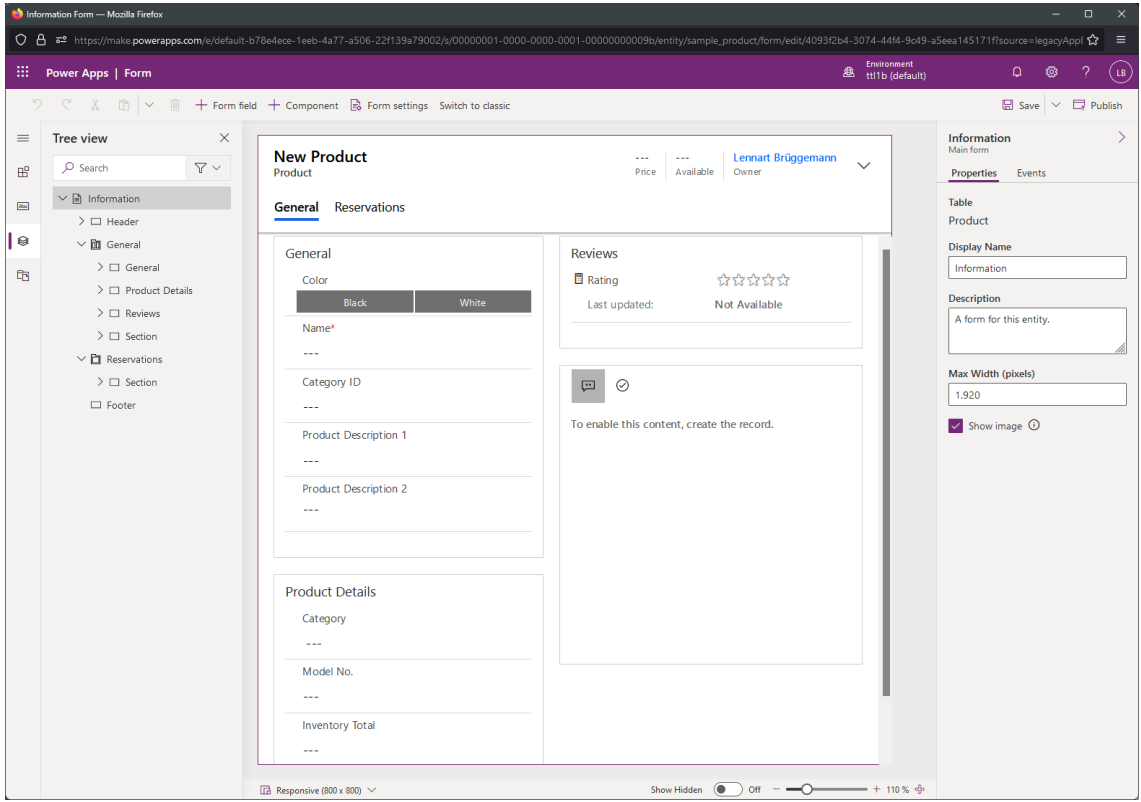

Figure 2.2: Design view of the *Asset Checkout* app

While security aspects, such as transport security (like HTTPS), are also relevant in locally running applications, they are more difficult to evaluate with PaaS models, as the responsibility for secure LCDP and LCA configuration lies with the vendor of the service. As Oltrogge et al. discovered [122, p. 10], every examined online service was vulnerable to at least one possible attack vector. Supporting and educating citizen developers in securely configuring their applications is thus of higher importance, when online-hosted LCDPs cannot be trusted with adequate security defaults. Security concerns and technical solution proposals will be further examined in chapter 4.

## 2.2 Types of low-code applications

Running a local LCDP can take two forms, depending on the architecture a platform uses. In the first case, described in section 2.2.1, the LCDP is both the development environment, where the LCA is designed and configured, and the runtime-environment where the LCA is executed. The other way is to host the LCA in a dedicated *client* application, that is separate from the development environment. In this case, the LCA is developed in the *main* application and is then deployed to the client application. The client application can either run on the same or other computers, or on additional devices, such as smartphones. This case is discussed in section 2.2.2.

### 2.2.1   Hosted in the same application

Running the LCA inside a host has certain advantages. For one, it simplifies deployment, especially when it's planned to not only build a single one-off, but multiple applications. Only the host LCDP has to be installed and is used for development and as execution environment. The LCA doesn't have to be installed separately (with possible additional dependencies), and multiple LCA can be run in the same environment. One example of this is the *SCOPELAND Direct Desk*, which supports this case in *interpreter mode* [120, p. 9], and is introduced in detail in section 2.4.

The disadvantage of this approach is that it circumvents the native distribution channels of the operating system the application will run on. For example, if a company administrates a fleet of mobile devices remotely, for example via mobile device management (MDM), it would be more difficult to distribute a LCA through mobile app stores (such as the *Google Play Store* or Apple's *App Store*), since only the host app could be installed on the devices. If the host app does not support provisioning of the apps it should run, it is more difficult to install additional LCAs remotely. If a LCA was available as a generated application (see subsection 2.2.3), the executable or application package (for example APK files[1] in Android) could be distributed as any other app on the platform. This issue is not limited to mobile devices, as remote management of client devices is also common for regular computers in large networks, for example using software like Microsoft Endpoint Manager [110].

### 2.2.2   Hosted in a separate application

A derivative of running LCA locally within the host LCDP is running it in an additional application (*client*) outside the host. This is illustrated by Microsoft's *Power Apps* application, that runs on Windows, Android, and iOS [29]. The LCAs themselves are developed using a browser-based tool (detailed in section 2.1.2), stored online and can then be downloaded into the *Power Apps* app on the desired target platform. Figure 2.3 shows the *Asset Checkout* sample application running inside *Power Apps* client on Windows 11.

Running the LCA in a separate application has the advantage that not the full LCDP has to be installed on the target device, only a smaller client program. Since in this example, *Power Apps* does not offer an on-premises installation, it's difficult to compare the space and hardware requirements. According to the Windows Store page for the *Power Apps* app [98], it requires about 140 MB of hard disk space and a device with 1 GiB of RAM — likely less than the full development environment would require. A smaller memory footprint is also beneficial, especially on mobile devices, as installation of large apps is undesirable from a user's perspective [13].

---

[1] An *Android package* contains the compiled application, resources, and metadata to distribute and install an app on the device — https://developer.android.com/guide/components/fundamentals

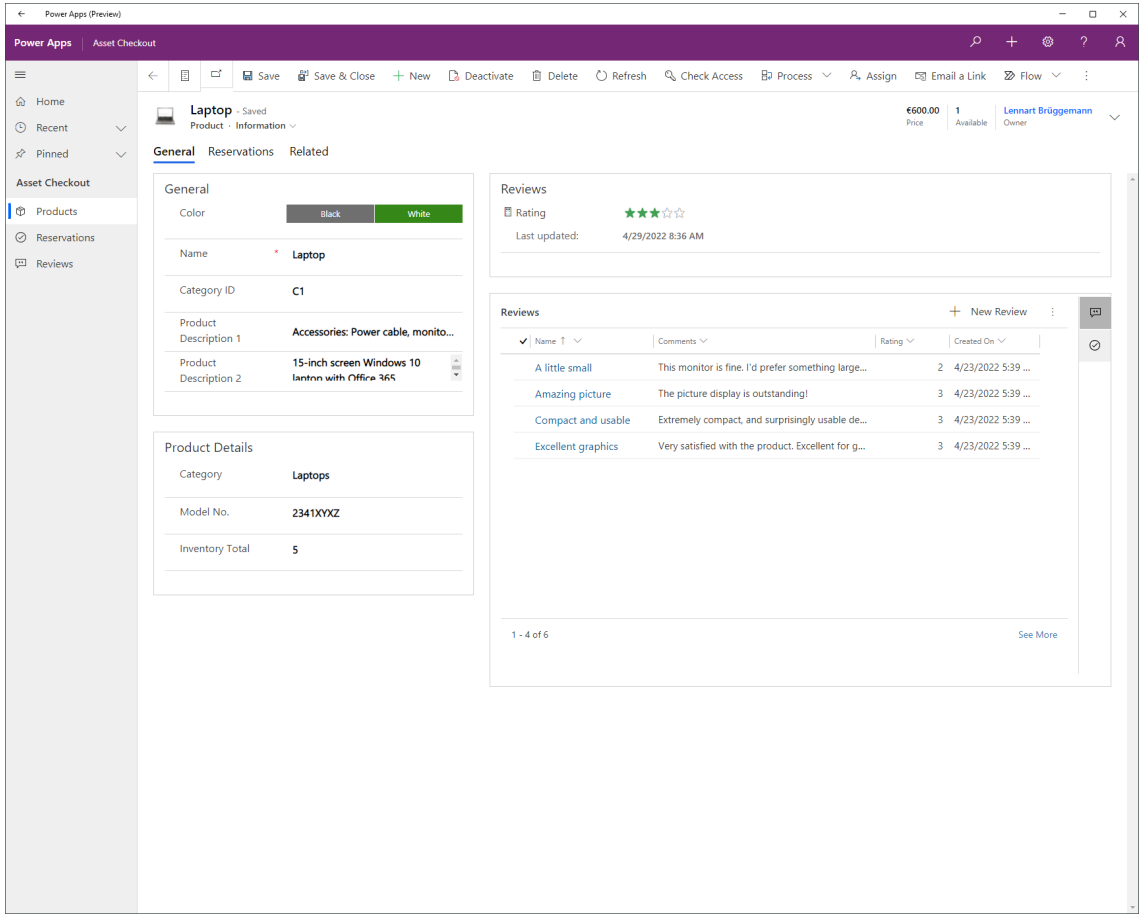

Figure 2.3: Power Apps *Asset Checkout* on Windows 11

This approach also offsets some disadvantages of installing the full host application on a device mentioned in subsection 2.2.1. While the LCAs themselves still aren't distributed through the operating system's native channels, the client takes care of listing and downloading the apps available online.

### 2.2.3 As generated application

Natively generated applications are the last possible form of LCAs. The LCDP translates the LCA into an application that runs on the target platform as any other conventionally developed program. Examples for this are Mendix, which can generate native Android and iOS apps using React Native [78], or SCOPELAND, which supports building both desktop and web applications (presented in detail in section 2.4).

The general advantage of generated LCAs is that they can be deployed and run independently of either the LCDP or a separate client. This minimizes the cost of distributing the LCA, if previous processes, such as technical infrastructure and existing knowledge, from existing (non low-code) applications are already in place and can be quickly adapted for new programs. This independence is also an important aspect for future maintainability: since generating a native application requires source code in some form, this source code can potentially be used and further developed without the original LCDP, for instance, should the vendor stop supporting their product later on. However, altering the generated source code has the drawback that changes to it cannot be ported back into the original LCDP, since that would require essentially a reverse-interpretation of code into the internal format of the LCDP.

The source code can also be the starting point for more traditional software-engineering methods. Static program analysis, gathering information about an application without executing it [32], can for instance be used to check the source code for possible inadequate function calls or parameters, such as insecure methods (for example without or outdated encryption) to connect to remote web servers. This is especially important if additional scripting code, handwritten by the citizen developers, was generated into the LCA. While even the automatically generated code cannot be expected to be free of security issues [122], code written by users without formal software development training should not be trusted by default either.

## 2.3  Extension of low-code applications

As their name implies, low-code applications can still contain a certain amount of custom code written by the users. Both hosted and generated applications can be extended that way. In SCOPELAND, these extensions are called *scripts* and are used to implement complex logic, that cannot be expressed with available functionality, or to interface with third-party software [120, p. 68]. In generated applications, the additional code is automatically embedded into the resulting code created by the LCDP [120, p. 47].

Suppose that a user wants to execute additional logic when a button in their application is clicked. The user doesn't need to know how exactly the *button clicked* event has to be implemented, the code generator produces the necessary scaffolding code and inserts the additional code written by the user into the method body of the event function. Listing 2.1 shows a simple example how such scaffolding code could look in WPF (Windows Presentation Foundation). Mendix uses a similar process when generating custom Java actions [83].

Unless there are additional checks in place that verify the custom code inserted into the application, this presents possible security risks. For example, a user could leave

```
1  public partial class SampleWindow : Window
2  {
3      public SampleWindow()
4      {
5          sampleButton.Click += OnSampleButtonClicked
6      }
7
8      private void OnSampleButtonClicked(object sender, RoutedEventArgs e)
9      {
10         // User code gets inserted here
11     }
12 }
```

Listing 2.1: Example of scaffolding code for a click event on a button in WPF

debugging statements, which print unsanitized user data into a publicly accessible log, in the application. A malicious developer could try to break out of the application by reading foreign files off the hard disk, or send sensitive data to a remote web service. Security analysis of user-written script code in generated applications and approaches to run code isolated will be discussed further in chapter 4.

While adding some small code fragments to the application may not be a problem for an experienced user, a citizen developer without programming training might struggle even with simple expressions. Supporting citizen developers in writing code, for example by providing specifically customized APIs or visual clues in the editor, is therefore also a responsibility of the LCDPs. A framework for the development of visual augmentations for a code editor and other methods to support developers are introduced in chapter 3.

## 2.4 SCOPELAND

SCOPELAND is a low-code platform made by the German company Scopeland Technology GmbH[2]. It is primarily aimed at building applications from new or existing databases, that is, the database is the primary element, and the application is built around it [120, p. 12]. In addition to the database that contains the primary data for the application, a second *meta-database* is individually created. The meta-database contains supplemental information [120, p. 32], such as allowed ranges for values, calculated columns or plausibility rules. This also enables the assignment of logical names to tables and columns, so user-friendly names, independent of length or character set constraints, can be used. Citizen developers might not be familiar with the technical names used internally to describe the database or the limitations imposed by the database system. For example, a column could have the logical name *Supervisor place of residence*, which would require special handling (such as explicit quoting due to spaces, escaping of special characters, or consideration of name length limits) if used as a physical name. Logical names can also be used for localization, since they can be translated to support multiple application languages independent of the physical names from the database.

---
[2] www.scopeland.com

SCOPELANDs primary development environment is called the *SCOPELAND Direct Desk* [120, p. 9], a Windows program used to build, configure and generate the applications. After a table is described with metadata, it can simply be opened and displayed, including the main columns of tables referenced via foreign keys [120, p. 33]. Figure 2.4 shows such a table opened in Direct Desk. In this case, the referenced *sector* and *country* columns are already resolved to the concrete values their foreign keys refer to.

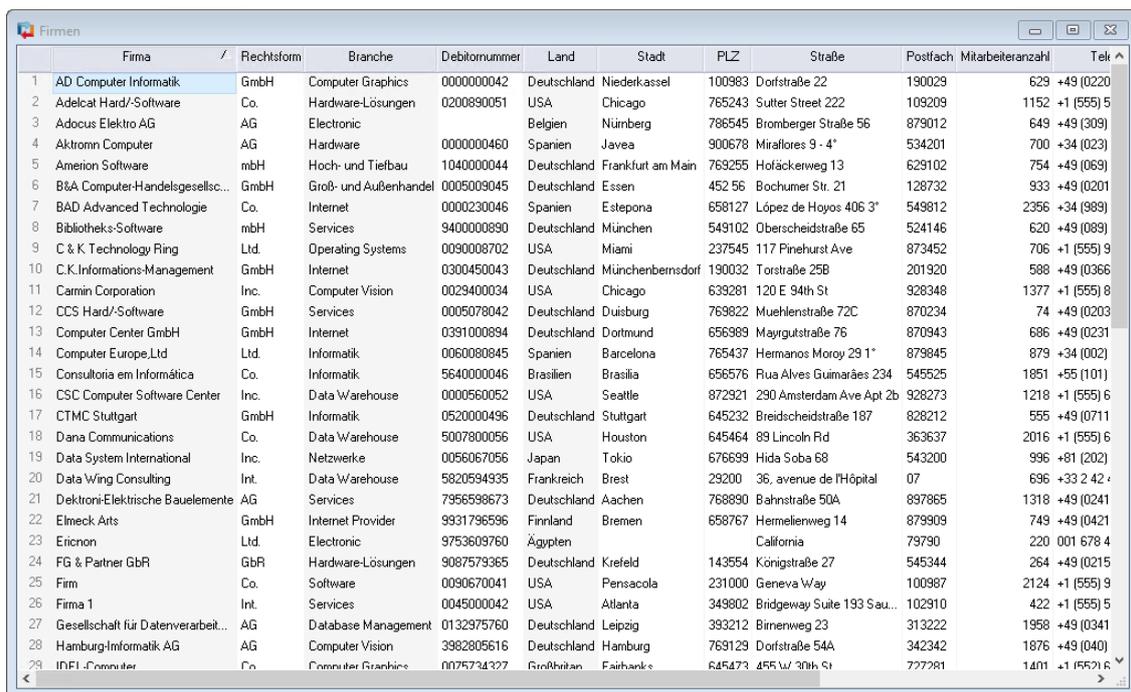

Figure 2.4: A table opened in SCOPELAND Direct Desk

Based on this table, a simple application can already be created. For example, displaying the currently selected data set in a window, with a UI automatically derived from the table's metadata and enabled editing capabilities, can be achieved by selecting the table to be displayed as a *form*. As shown in figure 2.5a, the *sector* and *country* fields are displayed as a drop-down box, which contain only the values available in their respective table. The data in form can be customized, for example, to display only certain columns or apply custom logic, such as sorting. Visually, all elements can be rearranged freely and formatted (such as fonts and colours).

Figures 2.4 and 2.5a show examples of an application that runs locally in a host environment, as introduced in subsection 2.2.1. Hosting an application in a separate application (subsection 2.2.2) is not supported in SCOPELAND; however, it is possible to generate a native application as described in subsection 2.2.3. SCOPELAND supports multiple target platforms, such as WPF [140, p. 4], ASP.NET or Java Server Faces (JSF) [120, p. 9]. Figures 2.5b, 2.5c and 2.5d show generated versions for each of the target platforms of the application originally hosted inside the Direct Desk (figure 2.5a). While the WPF version can be told apart by the native Windows title bar and control designs, the ASP.NET and JSF version running in a browser are virtually indistinguishable from each other.

Code generation for native applications is internally facilitated by exporting the representation of the hosted version of the application into an XML-based format called XDAML [120, p. 56]. The XDAML file contains no platform or database-specific code. The code generator for the target platform then translates the XDAML file into source code and uses the platform's development toolchain (for example Visual Studio [139]) to compile a native application. The availability of the source code also means there is less vendor lock-in, as at least the source for the generated application can also be edited and run outside SCOPELAND [141].

(a) SCOPELAND Direct Desk
Hosted, same application

(b) WPF
Generated, native Windows application

(c) ASP.NET
Generated, native web application

(d) JSF
Generated, native web application

Figure 2.5: Examples of hosted and generated SCOPELAND applications

# Chapter 3

# Supporting citizen developers in writing code

Ungar et al. [151] describe the concept of *immediacy* in programming, that is, a perceived connectedness of the programmer to the development environment — similar to how a race car driver can *feel* the connection of their finely tuned vehicle to the track surface, just by subtle changes experienced through the steering wheel. They distinguish between three types of immediacy: temporal, spatial and semantic. Temporal immediacy describes the delay between an event and its effect, for example, how quickly a programming environment can highlight an error made by the developer. Spatial immediacy refers to the physical distance, how close the visual notification of an error is to the faulty line of code. Semantic immediacy is the mental effort required to connect an error message to its cause. Their article concludes with an important point, that is not only relevant to professional software developers, but can be applied to anyone using code editors:

> ' *Attaining a reasonable level of immediacy is a precondition for effective debugging and can serve as a useful guide to the design of all aspects of programming environments.* '

This chapter will present existing research on code editors and APIs, and how they can be applied to improve the development experience when extending low-code applications with additional code. As part of this thesis, several of these ideas were used to build a programming framework that utilizes the code editor AvalonEdit [1] and allows fast and modular implementation of extensions that can help citizen developers in writing code for LCDPs.

---

[1] http://avalonedit.net

## 3.1  Static code analysis

As discussed before, the code written by citizen developers will either become part of the generated source code of an LCA or stored and run as part of a hosted LCA. This enables the use of static code analysis, an automated evaluation of the information exposed by the source code of an application, without actively executing it. A more formal definition was made by Prähofer et al. [128]:

> ' *Static code analysis works by analyzing the static structure and the elements of a program without actually executing it. It is therefore usually based on the source code of a program or an intermediate representation thereof.* '

This contrasts with *dynamic code analysis*, which gathers information on programs during their runtime, or, as defined by Binkley [14], '*takes program input into account*'.

Using static code analysis to find problems in code is not a new concept. In 1969, the group for the standardization of COBOL in the US Navy, led by Grace Hopper, developed a COBOL preprocessor called *NAVPREPILE-C*. It ran the code of a program against a set of validations, such as checking for syntax errors or non-standard statements, to reduce the time required for debugging and recompilation [16, p. 124]. A more recent example, which lends its name to tools until today [2], is *lint*, a program to find defects in C source code files, which was published in 1978 [66]. Today, analysers of these kind are common, both directly integrated into IDEs (for example Roslyn [93]) and as separate programs (such as FindBugs, a static code checker for Java [73]).

Static code analysis has multiple applications; it can not only be used to find errors in code, but also to check statements for problematic method calls (such as the use of insecure cryptography algorithms [161]) or to enforce a specific code style.

A simple example is the concatenation of two strings using the `String.Concat()` function in C#, which has the signature shown in listing 3.1.

```
public static string Concat (string str0, string str1);
```

Listing 3.1: Signature of the `String.Concat()` method overload for two strings in C#

The function takes two strings (`str0` and `str1`) and returns a new string containing the result of the concatenation (as, just like in Java, strings in C# are immutable). In listing 3.2, this function is invoked, but the return value is ignored, as a user could erroneously assume that the concatenation is performed directly on one of the input variables. Writing this code in Visual Studio will cause a suggestion from the static code analysis (that runs per default in the background) to appear, that looks similar to the one in figure 3.1.

---

[2] For example ESLint, a linter for JavaScript — `https://eslint.org`

```
1  var str0 = "Hello";
2  var str1 = "World";
3  string.Concat(str0, str1);
```

Listing 3.2: Ignoring the return value of the `String.Concat()` function

Besides the hint itself, messages like this often also include some tips on how to resolve the issue. Code analysis rules are often assigned a unique identifier (CA1806[3] in this case) combined with a link to a website with additional information.

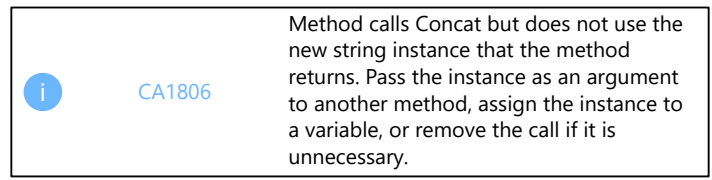

Figure 3.1: Replica of a Visual Studio static code analysis suggestion, based on [86]

Static code analysis can also detect errors that make the code not compilable without actually having to compile the code. Syntax errors, such as missing semicolons or misspelled (and not existing) functions can be detected during development time, before running the compiler and having to wait for error messages. As mentioned before, potential security issues can also be detected, for example code that is prone to SQL injection[4][87] or uses broken cryptography algorithms such as DES [88].

Implementing custom static code analysis rules is possible and detailed in section 3.5.

## 3.2 Mechanisms in code editors

Both simple and advanced code editors usually have some basic visual aids in common. *nano*[5], a minimalistic text editor for the command line and displayed in figure 3.2a, supports line numbers and syntax highlighting for a selected number of languages (which can be extended). Multipurpose text editors like *Notepad++*[6] may support more features. For example, *code folding* (the red ⊟ symbol on the left side), which collapses a certain region (like the body of a function) to temporarily reduce its used space, or highlighting of the currently selected line as shown in 3.2b.

More advanced editors as part of IDEs can include additional visual hints from external sources, such as code analysis. Figure 3.2c shows an example from Visual Studio 2022,

---

[3] '*Do not ignore method results*' [86]

[4] Specially prepared strings can be used to inject code into unchecked SQL statements to enable attacks on the underlying database, such as executing arbitrary code or accessing private data [149]

[5] `https://www.nano-editor.org`

[6] `https://notepad-plus-plus.org`

(a) nano, text editor for the command line

(b) Notepad++, a text editor for Windows

(c) Visual Studio 2022 IDE

Figure 3.2: Visual features in different code editors

which adds numerous extra indicators:

- In line 1, an *inline parameter name hint* is displayed, which shows the name of the method parameter from line 3.

- Line 4 displays a warning from static code analysis directly at the affected place (*inline diagnostics*, an example for spatial immediacy), the relevant code is also underlined.

- The opening and closing braces in lines 4 and 9 are connected by a dashed vertical line, indicating their pairing.

- The `return` keyword in line 6 is colourized with the same background colour as the type of the return value (`int`) in line 3.

- To the right of the line numbers is a vertical indicator showing saved (green) and unsaved changes (pale yellow) in the file.

- Line 8 shows an indicator for a recursive call on the far left, and a small icon of a screwdriver right of the line numbers signalling an available refactoring action.

Sulír et al. call these kinds of indicators '*visual augmentations*' and present an extensive taxonomy [146] with 29 different attributes distributed among 7 dimensions based on the study of over a hundred different tools for code editor augmentations. Taking the *inline parameter hint* previously shown as an example, it would be classified as an augmentation based on *code*, representing a *string* and displaying *text* directly *in the code*, at a certain *character*, which can be *not* interacted with and is implemented in an *existing* IDE.

The amount of possible combinations of all attributes is immense, and presents an opportunity to exactly tailor a particular augmentation to an existing use case in a low-code environment. This thesis proposes a framework that enables the quick development of custom augmentations for a code editor to support citizen developers when writing additional code for a LCA. As concrete requirements will vary based on the experience of users and the scope of the code added to the LCA, the framework will provide extensibility points, that allow for a wide amount of different visualizations to be developed.

## 3.3 Visual Augmentation Framework

The taxonomy of augmentations for source code editors by Sulír et al. [146] shows the many possibilities for additional visual indicators that can support the user. As shown in section 3.2, even minimalistic editors support basic features like syntax highlighting and line numbers. Advanced IDEs can supply more in-depth information about the code, such as context-dependent content depending on code analysis or visual indicators for specific code elements beyond syntax highlighting. However, these are limited to the IDE itself and are unavailable in embedded editors that a LCDP might use. Extending such an editor is an option to provide a citizen developer with additional information, guide him towards the correct usage of certain APIs or show interactive help texts.

Embeddable source code editor controls are available for many programming frameworks. Examples are the Scintilla [7] editor for C++ (with ports available for other languages, such as Java and C#) or RoslynPad [8], a cross-platform C# editor based on AvalonEdit, that can also be embedded in other applications.

The framework introduced in the following sections utilizes the extensible rendering process of AvalonEdit [57] to quickly create custom visual augmentations, that can be tailored to the specific use-cases a LCDP wants to help its users with. AvalonEdit was chosen as the foundation for implementing the framework, since it is free, open-source, and provides extension points to customize the rendering process for the text display (detailed in section 3.3.1). Furthermore, as of the time of writing, it is still actively maintained, while other projects, such as a .NET port of Scintilla[9], have not received recent updates or maintenance. Other available code editors for .NET, including AvaloniaEdit[10] or RoslynPad already use AvalonEdit as basis, and it is widely used with over 4.6 million downloads [121] to date.

While the framework aims to support as many of the dimensions and attributes defined by Sulír et al. as possible, there are certain limitations, mainly due to the nature of the framework (an extension to an existing editor) and the lack of usefulness of several possible combinations of attributes. For example, the dimension *IDE* is limited to *existing*, as the framework was developed with a specific editor in mind and cannot be used as an external tool. The attributes are not arbitrarily combinable, either because they are mutually exclusive (mainly in the *target* and *type* dimensions), or because there is no way to access them (except *code*, everything from the *source* dimension). There are also some limitations that stem from the underlying UI framework (WPF). While the *visualization* can be theoretically any UI element (from simple text strings to complex diagram controls), not all of them support all *interactions*, such as tooltips or interactivity in general.

---

[7] https://www.scintilla.org
[8] https://roslynpad.net
[9] https://github.com/jacobslusser/ScintillaNET
[10] https://github.com/AvaloniaUI/AvaloniaEdit

### 3.3.1 Overview of the AvalonEdit architecture

As the frameworks builds on the extensibility points of AvalonEdit, a short overview of its architecture based on [57] will be given. There are three different kinds of objects that can be added to extend the editor: classes derived from `VisualLineElementGenerator` and implementations of the interfaces `IVisualLineTransformer` and `IBackgroundRenderer`. Each instance of an editor has a list for each particular type that will get evaluated during the rendering process.

Figure 3.3 shows the transformation sequence that is used to render text from the initial internal structure in the editor to the output on the screen. An instance of the `DocumentLine` class contains the plain text of a single line as displayed in the editor. At that stage, the text can be modified by inserting additional elements into the text. These can be more text strings, hyperlinks or any custom UI element, from a simple button to more complex layout panels containing multiple other controls. This is enabled by instances of the abstract `VisualLineElementGenerator` class. Derivatives have to implement two methods that will be called automatically for each line during the rendering process: The first is `int GetFirstInterestedOffset(int startOffset)`, which has to return either the offset in text the generator wants to insert an element, or the value `-1` if the generator will not modify the text in that particular line. The other is `VisualLineElement ConstructElement(int offset)`, which enables the construction of a `VisualLineElement` (which contains the UI element to be added) that gets inserted at the supplied offset.

At the next stage, a `VisualLine` is built by the editor, and implementations of `IVisualLineTransformer` will have their `void ColorizeLine(DocumentLine line)` method called, which can be used to alter visual elements in the particular `VisualLine`, such as changing fore- and background colours or setting font-related properties. This can, for example, be used to implement syntax highlighting by colouring specific keywords in the text. The result is a transformed `VisualLine`, which is then processed into one or more `TextLines`, which are used to calculate the flow of the text. As the content (and length) of a line might have changed, it is required to fit the text into the space available to the editor to use features such as word-wrapping (if enabled). This is handled by AvalonEdit internally, which uses existing classes from the `System.Windows.Media.TextFormatting`

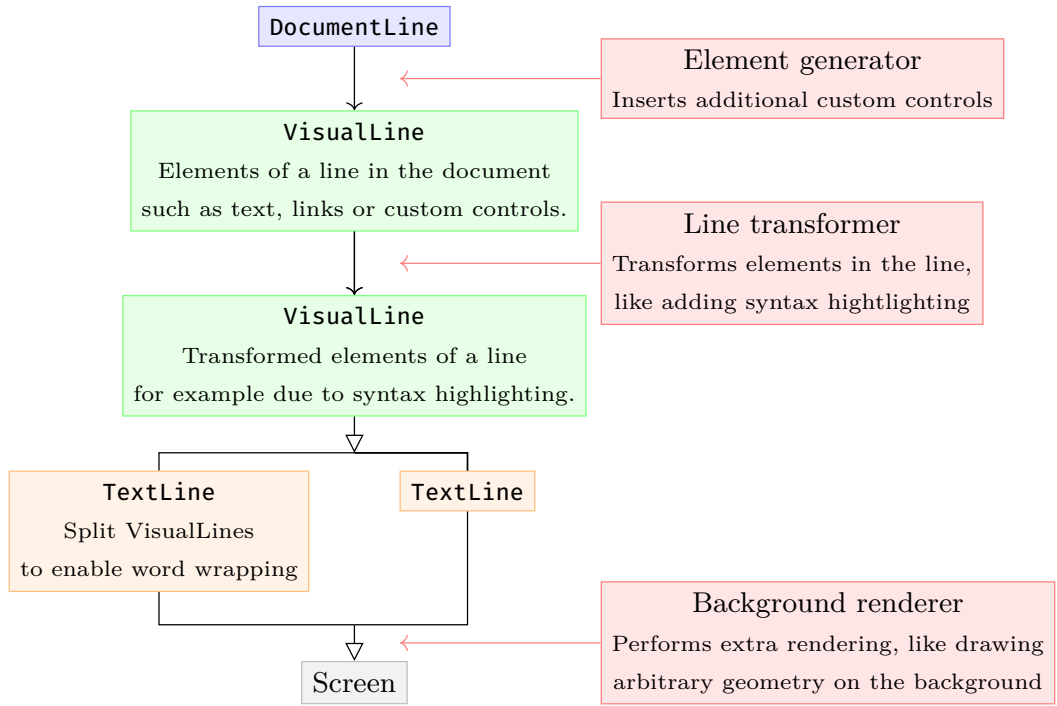

Figure 3.3: Rendering pipeline of AvalonEdit. Adapted from [56]

namespace [105] for that task.

The final extension point is the `IBackgroundRenderer` interface. Implementations can use the `void Draw(TextView textView, DrawingContext drawingContext)` method to draw arbitrary visual content on various layers (such as the background area, or the layer containing the blinking caret). For example, indicators for errors or warnings (like underlining text with squiggly lines) can be implemented that way.

### 3.3.2 Implementation

The augmentation framework was developed by the author of this thesis, its source code is publicly available on GitHub [18]. This section explains the core concepts and presents two examples on how the framework can be used.

The three extension points introduced in section 3.3.1 are the basis used for the implementation of the framework. An augmentation is represented by an object of the `Augmentation` class, which manages instances of the three extension objects internally. Each augmentation is responsible to transform a specific part of the displayed text, which is determined by assigning one or more fixed strings, or one or more regular expressions that match specific sequences of characters or words in the text. To determine what kind of transformation should be applied to the text, a simple fluent API to specify colours, font settings or interactive elements was developed. Internally, these API calls build and configure objects of the three extension classes and add it to the instance of the `Augmentation` class that gets currently configured. For demonstration purposes, augment-ations to provide configurable syntax-highlighting for Smalltalk code were implemented. Smalltalk was chosen for its concise syntax, the sample '*Smalltalk syntax in a postcard*' [156], shown as part of figure 3.4, shows most of Smalltalk's keywords and language constructs in a single method and lends itself to show many augmentation styles at once.

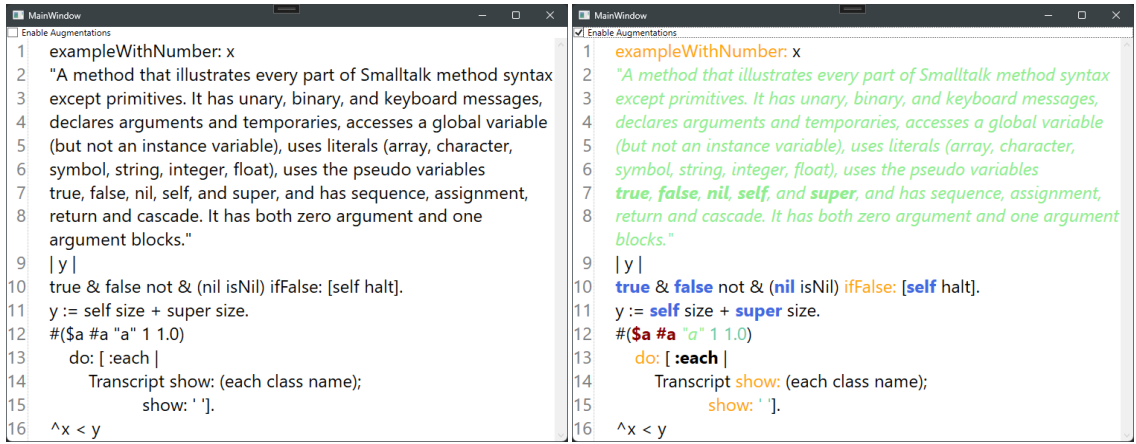

(a) Smalltalk code without augmentations    (b) Smalltalk code with augmentations

Figure 3.4: Syntax highlighting augmentation for Smalltalk code

The implementation, shown in listing 3.3, consists of six augmentations for each of the syntax elements to be decorated. A regular expression for a particular syntax element was defined and set for each of the augmentation. Afterwards, different styles, such as foreground colours or font options, were added. The result is shown in figures 3.4a (the plain text without syntax highlighting) and 3.4b (with augmentations applied). For example, comments are shown in light green colour with italicized font style and keywords are displayed in blue and with bold font weight. This example uses exclusively the `IVisualLineTransformer` extension point from AvalonEdit.

An example utilizing both the `VisualLineElementGenerator` class and the `IBackgroundRenderer` interface is shown in figure 3.5. In this case, a citizen developer may have used a hashing function using the SHA-1 algorithm. As SHA-1 is no longer considered secure [69], an augmentation was developed to warn the user upon usage by underlining the method call and displaying a tooltip showing a text that suggests alternatives.

Figure 3.5: Augmentation warning of the use of a SHA-1 method

The code required to implement such an augmentation is shown in 3.4 and consists of only five lines, defining the regular expression used to match the function call, the text for

```csharp
public static class SmalltalkHighlighting
{
    public static Augmentation[] GetAugmentations(TextView textView)
    {
        var parameterRegex = new Regex(@":\w+");
        var symbolRegex = new Regex(@"[$#]\w+");
        var numberRegex = new Regex(@"\b\d+(\.\d+)?\b");
        var messagesRegex = new Regex(@"\w+:");
        var commentRegex = new Regex("\"(.|\r|\n)*?\"");
        var stringRegex = new Regex(@"\'((.|\r|\n)*?)\'");
        var keywordsRegex = new Regex(@"\b(self|super|true|false|nil)\b");

        var comments = new Augmentation(textView).ForText(commentRegex)
            .WithForeground(Brushes.LightGreen).WithFontStyle(FontStyles.Italic);

        var keywords = new Augmentation(textView).ForText(keywordsRegex)
            .WithFontWeight(FontWeights.Bold).WithForeground(Brushes.RoyalBlue);

        var messages = new Augmentation(textView).ForText(messagesRegex)
            .WithForeground(Brushes.Orange);

        var numbersAndStrings = new Augmentation(textView).ForText(numberRegex, stringRegex)
            .WithForeground(Brushes.MediumAquamarine);

        var symbols = new Augmentation(textView).ForText(symbolRegex)
            .WithFontWeight(FontWeights.Bold).WithForeground(Brushes.DarkRed);

        var parameters = new Augmentation(textView).ForText(parameterRegex)
            .WithFontWeight(FontWeights.Bold);

        return new[] { keywords, messages, numbersAndStrings, symbols, parameters, comments };
    }
}
```

Listing 3.3: Implementation of the `SmalltalkHighlighting` class

the tooltip, and the definition of a `Geometry`[11] for the decoration. Here, a pre-configured decoration in the shape of a horizontal square bracket is used. Custom geometry can be injected by using a delegate method that takes a `Rect` as input to define the bounds of the decoration and returns an implementation of the `Geometry` class. The decoration is drawn by an implementation of `IBackgroundRenderer` on the background layer of the editor area using the coordinates from the bounding box of the text to adorn. The tooltip isn't part of the drawing, but a property of a `TextBlock`[12], that is inserted as an inline control (using a `VisualLineElementGenerator`) and replaces the plain text previously at its offset in the document.

---

[11] An abstract class to define the geometry of arbitrary 2D-shapes in WPF — https://docs.microsoft.com/en-us/dotnet/api/system.windows.media.geometry

```
1   var underlineAugmentation = new Augmentation(Editor.TextArea.TextView)
2       .WithDecoration(UnderlineBracket.Geometry)
3       .WithDecorationColor(Brushes.Red)
4       .WithTooltip("SHA1 is cryptographically broken, please use a currently secure function like
    ↪ SHA-512.")
5       .ForText(new Regex(@"\.HashWithSha1(.*)?\)"));
```

Listing 3.4: Implementation of the SHA-1 warning augmentation

Figure 3.6 shows a high-level overview of the augmentation framework. Classes and interfaces that are part of AvalonEdit are coloured in green, implementations of these and additional classes that were developed for the framework are blue. An `Augmentation` is the core component of the framework. The class manages lists of each individual extension point implementations, and has access to an instance of AvalonEdit's `TextArea` class, which exposes properties to add and remove elements to its rendering pipeline. The `AugmentationExtensions` class contains the methods used to configure each `Augmentation` in a fluent syntax style, shown in listing 3.4, for example.

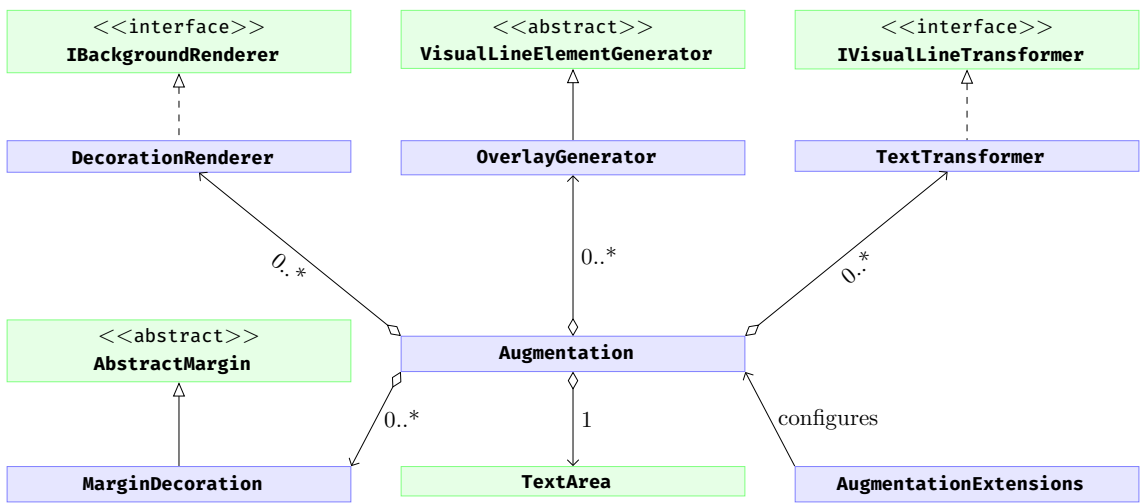

Figure 3.6: Augmentation framework class diagram

---

[12] A simple WPF control to display formatted text — https://docs.microsoft.com/en-us/dotnet/api/system.windows.controls.textblock

### 3.3.3 Applications

While syntax highlighting or text decorations for source code are nothing new, the idea behind the framework is to provide augmentations for specific use-cases that are not common enough to be supported by available editors by default. Although the framework was developed in the context of LCDP, its usage is not limited to them — any interested user or organization using AvalonEdit can utilize it to develop augmentations. A concrete example for an application of the framework is the *clear text display* and *number display* mode in SCOPELAND [120, p. 58]. Every table and field in SCOPELAND has a unique numeric identifier with a prefix of the object it refers to (like `T1000` for a table), which can be used to express relations between tables. Figure 3.7 shows a simplified database relation between a product that is assigned a single category. If the user wanted to refer to the category of a product in table `T1000` for use in a script or as part of a condition, they would use the expression `F1001.T2000.F2001`. In *number display* mode, this is both unwieldy and hard to understand. The alternative *clear text display* mode, however, would show `Category ID.Categories.Name` in this case, which is easier to read.

| Products (T1000) | | |
|---|---|---|
| PK | Product ID | F1000 |
| FK | Category ID | F1001 |
|    | Name | F1002 |

| Categories (T2000) | | |
|---|---|---|
| PK | Category ID | F2000 |
|    | Name | F2001 |

Figure 3.7: Simple relation between products and categories

This mode can be easily implemented with the augmentation framework. Given a mapping of the fields of a table and the desired names, the names can be overlaid over the identifiers, as shown in figures 3.8a and 3.8b, with the source in listing 3.5.

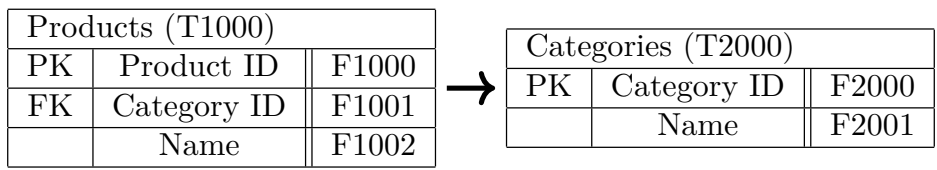

(a) Number display mode  (b) Clear text display mode

Figure 3.8: Two different display modes for identifiers in code

### 3.3.4 Discussion

The augmentation framework can be used to improve the different types of *immediacy*, as introduced at the beginning of this chapter. Augmentations fall primarily in the *spatial* categories. As all additional visual elements generated by the extensibility classes of

```csharp
public static IEnumerable<Augmentation> GetAugmentations(TextView textView)
{
    foreach (var (field, name) in GetFieldMapping())
    {
        yield return new Augmentation(textView)
            .ForText(field)
            .WithBackground(Brushes.DarkGray)
            .WithForeground(Brushes.White)
            .WithOverlay(name);
    }
}
```

Listing 3.5: Implementation of the augmentation for the clear text mode

AvalonEdit as part of the rendering process, they are available as soon as the editor has finished loading. *Temporal* immediacy is thus not a relevant part of the augmentations built with the framework, as their display is usually instantaneous upon activation, and the developer has little influence (short of explicitly adding delays) over their timing. Spatial immediacy is most easily achieved by the various colouring options provided by the framework, as they directly adorn the text strings they are applied to. The same applies to text decorations, such as the underlined method call from figure 3.5. *Semantic* immediacy largely depends on the developer adding concise information to the augmentations, such as tooltips or text replacements. Since the editor itself does not support further user interaction (such as debugging, or showing more diagnostics in tool windows or menus), there are limited opportunities to improve the display of semantically related information.

The dimensions and their attributes from the taxonomy of Sulír et al. [146] can now be compared against the possible augmentations build with the framework. The following sections give a short explanation of each dimension and how they can be implemented, the attributes of each dimension are denoted in *italics*.

**Source**

The source dimension describes the origin of the data that is represented by the augmentation. Searching for character sequences via simple comparison or regular expression is essentially static code analysis. However, since it is possible to pass delegates to an augmentation, it is theoretically possible to execute arbitrary functions which in turn could provide data from other sources. For example, a developer could write a function that passes the code written in the editor to an external library. If the library built and run the code, perform some kind of dynamic analysis, and return specific offsets in the input code that should be augmented, this would be a form of dynamic code analysis. While other attributes of the source dimension would not be impossible to implement, the primary source for the augmentations is thus *code*.

**Type**

The type denotes the possible states an augmentation can take, for example it can be simply displayed or not (*boolean*), or show an icon from a fixed collection of images (*fixed enumeration*). The types are largely dependent on the implemented augmentations themselves. If the developer creates multiple augmentations to implement syntax highlighting (as in figure 3.4b), the augmentations represent a *fixed* number of syntax elements (comments, symbols, messages and so on). The switch between numeric and clear text display from subsection 3.3.3 is an example of *variable enumeration*, since the possible field and table identifiers can come from an external source with the exact number of combinations unknown, or even determined during runtime as new fields and tables are created.

**Visualization**

The visualization dimension depicts in what way the augmentation is displayed. The framework can be used to implement all types of visualizations. *Colour* and *text* are for example shown in figure 3.8b. The underline *decoration* from figure 3.5 can be substituted for custom *icons* and *graphics*, as WPF supports both drawing generic shapes and images loaded from files.

**Location**

The location dimension describes the general position an augmentation is placed. *In code* is shown is most of the previous examples, such as figure 3.5. *Left* and *right* can be used to show text, glyphs, or images in the left margin of the editor or on the right end of a line, shown in figure 3.9.

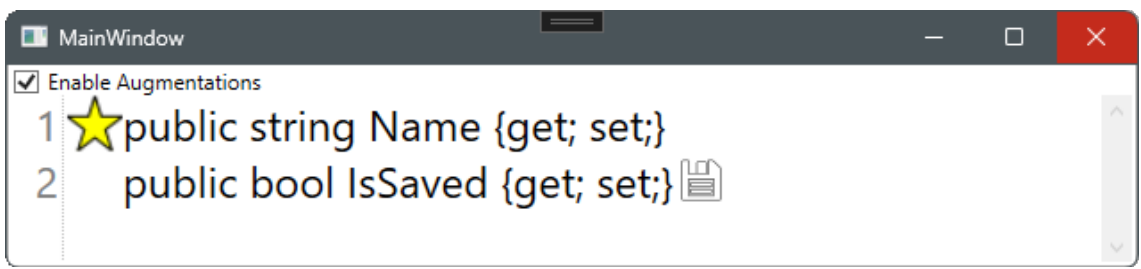

Figure 3.9: A star glyph and a floppy disk image shown to the left and right of the code area in the editor

**Target**

The positioning of the augmentation in the editor is defined by the target dimension. As the augmentations are generally based on offsets of the text in the editor, it is possible for an augmentation to span a single *line*, *line ranges*, *character ranges* or the whole *file*. The Smalltalk syntax highlighting example from figure 3.4b shows all options, except for an augmentation of the entire file, which can easily be implemented by a regular expression that matches every symbol in a file.

**Interaction**

The interaction dimension describes the possible forms of interaction with an augmentation. All available attributes can be implemented with the framework. *Popover* functionality is shown in figure 3.4, which displayed a tooltip when hovering with the mouse over an augmentation. It would be possible to display a more sophisticated tooltip, for example containing a link to a help page; thus *navigation* is also a viable option. In figure 3.8b several strings are replaced with text blocks to show them with different fore- and background colours. As any available control can be inserted in the code, which includes interactive controls like calendars or text boxes, the augmentations also support *changes*.

**IDE**

The IDE dimension denotes whether the augmentation is implemented for an existing IDE or editor, or as an external tool specifically created for the augmentation. As the framework build to extend AvalonEdit, it is implemented for an *existing* editor.

## 3.4 Improving access to APIs

### 3.4.1 Wrapping of existing APIs

The augmentation framework introduced in the previous chapter is essentially a wrapper over the existing extensibility points offered by AvalonEdit. It provides a simplified interface to apply colouring or decoration of text elements with only a few lines of code. Implementation of the necessary classes and logic required to build the augmentations is abstracted away from the user and exposed via an API that is specifically tailored to lower the barrier for creating them. A similar approach can be taken to offer an entry point for APIs that are either difficult to use correctly, or require special knowledge to prevent potential configuration problems, such as security issues.

Jugel et al. [67] categorize wrapper libraries as a domain specific language (DSL), '*which allows expressing the concepts of a specific domain that would be much harder to program*'. More specifically, they are a variation of an *internal* DSL, that, in comparison to an *external* DSL, does not consist of any new syntactic elements that would require a different development environment. Wrapper libraries exist to '*provide a mapping, refinement, or reduction of an underlying API's existing syntax*' [67], which makes them suitable for use in an API that aims primarily towards citizen developers.

For example, suppose a citizen developer has to write code to calculate the SHA512 hash of a string. This is an easy task, as the .NET API provides readily available functions to hash data with various hashing algorithms. However, usage of the API requires knowledge about reading a UTF-8 input string into a byte array, calling the correct function of the API and converting the resulting hash back into a string — something a citizen developer might not be acquainted with. A correct implementation may look similar to the one in listing 3.6.

```csharp
public static class HashFunctions
{
    /// <summary>
    /// Hashes a string using the SHA512 algorithm.
    /// </summary>
    /// <param name="input">An UTF8 string to hash.</param>
    /// <returns>
    /// The SHA512-hashed equivalent of <paramref name="input"/>,
    /// as a uppercase hex string without dashes.
    /// </returns>
    public static string HashAsSha512(this string input)
    {
        using (var sha512Hash = SHA512.Create())
        {
            var inputBytes = Encoding.UTF8.GetBytes(input);
            var hashedBytes = sha512Hash.ComputeHash(inputBytes);
            return Convert.ToHexString(hashedBytes);
        }
    }
}
```

Listing 3.6: Hashing a string with SHA512 in C#

For a developer not familiar with the API, there are a few aspects that are not immediately apparent. An obvious way would be to simply call the constructor of the `SHA512` class, which would fail as it is `protected` and only callable by derived classes. The official documentation for the `SHA512` class [103] offers an example using the `SHA512Managed` class, which is marked as obsolete [104] and only then points to the correct usage of the `Create()` method on its parent type. As its base `HashAlgorithm` class implements the `IDisposable` interface, it is also important to wrap calls to the `Create()` method in a `using` statement, to automatically release all internally allocated resources afterwards. Based on examples of improvable aspects of APIs [67, p. 359], specifically the omission of obsolete or unused classes and the difficulty to choose the correct functions, the `SHA512` class is a practical candidate for a simplified API that enables easier access for citizen developers.

The amount of work involved to design a wrapper library depends on the requirements of the citizen developers. A five-step process is proposed by Jugel et al. [67], which can be summarized as:

1. Import of an existing or creation of a new API-model, i.e., selection of an API to be wrapped.

2. Refinement of the model. Selection of classes and concepts that are of interest for the consumer, while avoiding existing usability problems of the API.

3. Improving the naming and annotations of the selected classes and methods. Technical names are replaced with more readable versions that can be better understood by the consumer. New classes and methods can optionally also be added.

4. Implementation of the actual code for the API. This can be either done manually, or with a special code generator, that was also developed as part of [67].

5. Testing of the resulting wrapper library. While the underlying code of the original library may have been fully tested previously, the new wrapper can still introduce errors or inconsistencies, and needs to be covered by new unit or integration tests.

Providing a library with cryptographic hashing functions to the citizen developer might thus start with a selection of supported algorithms. Functions that are prone to collisions, considered insecure or not applicable to the particular domain can be omitted, the API then only offers implementations that are to be used by the developers. The new functions could be implemented as extension methods, so they can be directly called on string variables and users don't need to create objects to use them. While the wrapped functions are part of .NET, tests should still be written for the created wrapper functions, to ensure that they perform as expected.

Adding support for the wrapper library to the editor is relatively easy. Compared to the previous sections, this example uses RoslynPad as an editor, since its built-in support for Roslyn enables the integration of external libraries, also called assemblies in .NET. Assuming that the library is referenced from the editor project, the assembly, in this case containing the namespace `WrapperApiSample.HashFunctions`, can be added to a new instance of the class `RoslynHost`, which is in turn passed to the editor during initialization. Listing 3.7 shows the relevant parts required for the integration of an additional library. When running the editor, the namespace for the desired class can be imported with a `using` directive, and RoslynPad will detect the used method (as implemented in listing 3.6) and display its documentation [13] as tooltip (figure 3.10).

---

[13] .NET uses a special XML syntax, informally called *XMLDoc*, to decorate code elements with documentation — https://docs.microsoft.com/en-us/dotnet/csharp/language-reference/xmldoc

```
1  RoslynCodeEditor editor = new RoslynCodeEditor();
2
3  var host = new RoslynHost(additionalAssemblies: new[]
4  {
5      Assembly.Load("RoslynPad.Roslyn.Windows"),
6      Assembly.Load("RoslynPad.Editor.Windows")
7  }, RoslynHostReferences.NamespaceDefault.With(assemblyReferences: new[]
8  {
9      typeof(object).Assembly,
10     typeof(WrapperApiSample.HashFunctions).Assembly,
11 }));
12
13 editor.Initialize(host, new ClassificationHighlightColors(), "C:\\WorkingDirectory",
       string.Empty);
```

Listing 3.7: Integration of an external library in RoslynPad

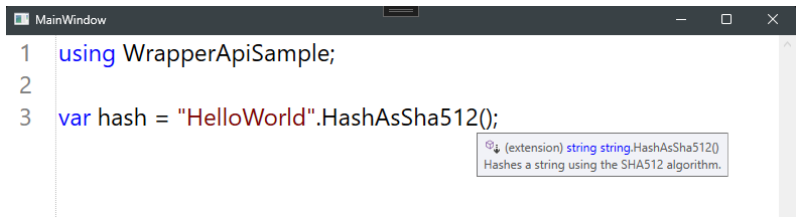

Figure 3.10: Calling the example wrapper API in RoslynPad

### 3.4.2 API Documentation

Well-structured documentation is also an important aspect for the correct use of APIs, as shown in a study by Gorski et al. [53]. They compared the effects of supplemental security information in official API documentation by adding code comments or a section specifically dedicated to a security feature. Four different variations were evaluated to find the areas the participants read most attentively and their effect on the results of specific programming tasks. They concluded that adding detailed information has a significant impact on the security of the resulting solutions, as none of the participants in the control group, using the standard documentation without supplemental help, managed to implement the security feature correctly. The sample group was composed of students with little programming experience, which shows that users without in-depth knowledge of software development (including citizen developers) profit from further documentation and produce more secure code.

Another study [54] showed that displaying advice upon usage of insecure cryptography functions in Python improved the security of the solutions implemented by the study group. The advice messages contained an explicit warning and offered the user meaningful options to continue: a secure action with a code example for a secure alternative implementation, and an insecure action with a hint how to suppress the warning. They conclude that most of the participants adhered to the advice and improved their formerly insecure code, however the warnings did not have a positive effect on the perceived usability of the API.

**Implementation of context-based advice**

The results of the two studies by Gorski et al. [53, 54] can be applied to develop context-based advice implemented with the augmentation framework introduced in section 3.3. As the concrete libraries or code expressions that should be covered by advisories in a LCDP can vary, the framework uses a generic approach to enable custom data sources to supply the information displayed. The classes containing the content of an advice implement the `IAdviceModel` interface, which includes information similar to the sample from Gorski et al. [54, Figure 1], such as code examples, secure and insecure actions, and hyperlinks to provide the citizen developer with additional resources. Implementations of `IAdviceModel` can come from any source, so it is possible to build advice models that are defined directly in code, or are supplied from a database or XML files.

Suppose a LCDP supports extension with a custom scripting language, which provides specific functions to insert newline characters into string expressions. Users may be accustomed to inserting the escaped newline sequence \n, which has a different meaning in that language. To inform the user to use the provided functions, an augmentation that detects and underlines usages of the newline sequences is developed. Similar to the security advice from Gorski et al. [54] that were printed in a terminal upon usage, the user can click the affected part of the code and is presented with a pop-up that contains details of the issue and is provided with sample code to address the problem, as shown in figure 3.11.

Compared to the implementation from [54], where the advice messages are part of the API, the implemented augmentation is an external component that is independent of the API it analyses. The advantage of this approach is that any APIs can be targeted with advice, and it is not necessary to modify the underlying existing libraries, which may not be possible (if the library is not open source) or not desired by its original authors. Advice can be tailored to target audiences and customized with appropriate messages, sample code and hyperlinks. Users can directly copy the sample code from the pop-up window into their editor, and click on the link at the bottom to open the target website in their browser. However, since the advice are implemented as part of an existing editor, they cannot be universally used elsewhere. They have to be deployed as part of a development environment, and, as they are not part of the API, they are also unavailable to developers not using that environment. Implementing advice as part of an API has the advantage that they are generally available to all consumers, as long as they are evaluated by the IDE that is used.

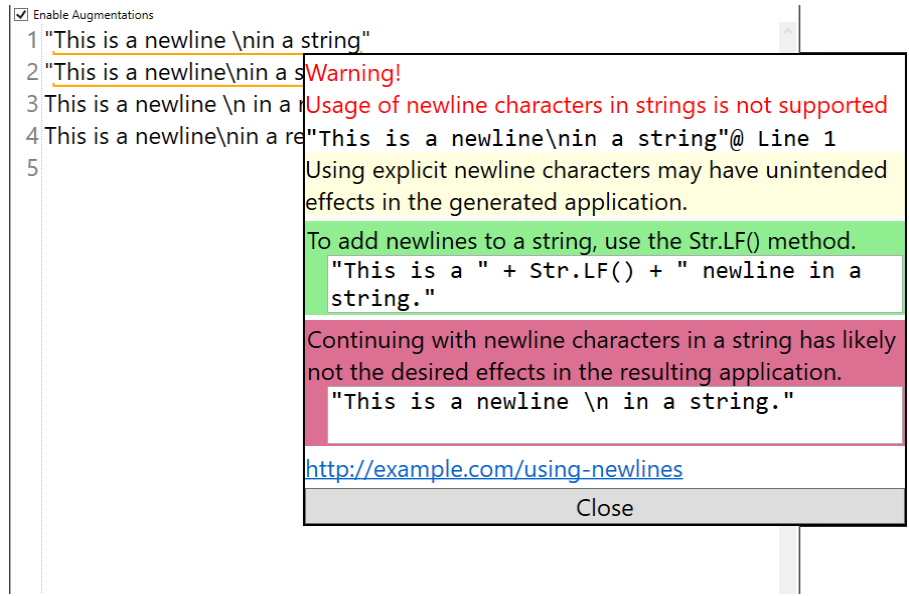

Figure 3.11: Advice augmentation triggered by the usage of an unsupported escape sequence.

## 3.5 Custom static code analyzers

The augmentation framework shown in section 3.3 relies on string comparisons and regular expression matching to identify the words or statements to adorn, and is essentially a simple form of syntax analysis. To distinguish, for example, whether a string is explicitly assigned to a variable, used as a function parameter, or is part of a comment, a more sophisticated analysis method is required. This section uses Roslyn, .NET's compiler platform, to implement code analysis rules and matching fixes that can be used to automatically resolve the issues found.

Microsoft provides the *.NET Compiler Platform SDK* [106] to interact with Roslyn and its syntactic and semantic code models. The software development kit (SDK) ships with the necessary tools to build custom code analysis rules called *analyzers*, to find, and automated corrections called *code fixes*, to address code issues. To ease development of analyzers, the SDK offers project templates for Visual Studio that provide the scaffolding and generate the basic infrastructure for analyzers, code fixes, packaging to distribute the analyzers, and unit testing projects.

To demonstrate how static code analysis can be used to support citizen developers, a Roslyn analyzer and code fix, that perform similar to the detection of unescaped newline sequences in section 3.4.2, were implemented for this thesis. As these analyzers are primarily developed for use in Visual Studio, there is no standard way to offer their functionality in external editors like RoslynPad used in the previous sections, thus limiting their usefulness for developers not using an IDE. This section also shows how the implemented analyzer can be integrated into RoslynPad, and enabling citizen developers to use them as part of an editor in their LCDP.

As with previous examples, the full source code is available on GitHub [18].

### 3.5.1 Implementation of an analyzer

The project templates already include the basic scaffolding code to build a simple analyzer, including the definition of a diagnostic rule, its localization and initialization. Therefore, it is only necessary to customize the predefined attributes of the rule, such as its severity and the messages the analyzer will emit (similar to the illustration in figure 3.1), and add the concrete analysis logic. Listing 3.8 shows the essential parts of an analyzer that detects the newline escape sequence (\n) in string literals.

```csharp
public class NewlineAnalyzer : DiagnosticAnalyzer
{
    private static readonly DiagnosticDescriptor Rule = new DiagnosticDescriptor("DiagnosticId",
        "Title", "Message", "Category", DiagnosticSeverity.Warning, true, "Description");

    public override void Initialize(AnalysisContext context)
    {
        context.RegisterSyntaxNodeAction(AnalyzeSyntaxNode,
            SyntaxKind.StringLiteralExpression);
    }

    private static void AnalyzeSyntaxNode(SyntaxNodeAnalysisContext context)
    {
        if (context.Node is not LiteralExpressionSyntax node) return;
        if (!node.Token.ValueText.Contains("\n")) return;

        var diagnostic = Diagnostic.Create(Rule, node.GetLocation(), node.Token.ValueText);
        context.ReportDiagnostic(diagnostic);

    }
}
```

Listing 3.8: Implementation of a Roslyn analyzer to detect unescaped newline sequences in string literals

The ID of the rule, along with its messages and severity, is defined in line 3. The `Initialize()` method is automatically called by the code analysis engine to set up and register the method containing the actual analysis logic, `AnalyzeSyntaxNode()`.

Source code is represented as a tree of syntactic elements in Roslyn. The SDK provides a syntax visualizer extension for Visual Studio that lets the developer inspect the state of the syntax tree in real time. Figure 3.12 shows how the structure of a simple string assignment to a variable that would trigger a diagnostic warning is structured. The *StringLiteralExpression* element that represents the text of the string itself is selected. A corresponding enum value is passed to the registration method in line 7 so that the analyzer is not called for the numerous other syntax elements.

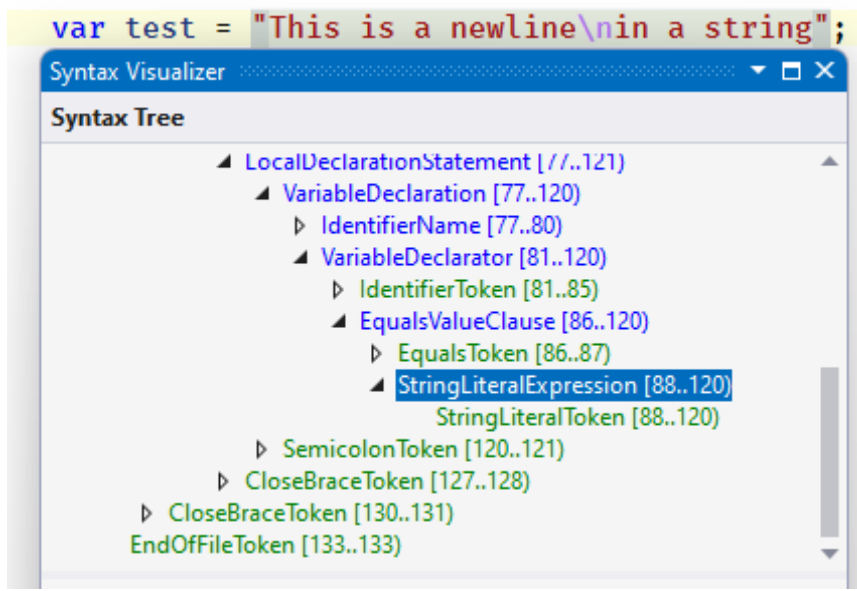

Figure 3.12: Roslyn's syntax visualizer showing the structure of a string literal variable assignment

If the check for a `\n` inside the string literal, shown in line 13, is successful, a new diagnostic is created. The diagnostic is supplied with the exact location and statement text of the analyzer violation, which enables the emitted message to be as precise as possible.

### 3.5.2 Implementation of a code fix

Analogous to the analyzer, the project templates already contain the basic infrastructure to implement an (optional) code fix that complements the analyzer. The implementation is similar, though the code fix is dependent on the diagnostic and receives the information reported by the analyzer from listing 3.8, line 16. Listing 3.9 shows how the code fix implementation inspects the string literal and creates replacement values that are applied to the code in the editor when the fix is invoked by the user. A new literal with the updated values is created, while the *trivia* (such as whitespace or comments [93]) is preserved to keep the string otherwise the same as before. The final step is to replace the old string literal in Roslyn's syntax tree. The practical application of this code fix is shown as part of the next section in figure 3.13.

```
1  private async Task<Document> ReplaceNewlineAsync(Document document, LiteralExpressionSyntax
   ↪ expressionSyntax, CancellationToken cancellationToken)
2  {
3      var token = expressionSyntax.Token;
4      var updatedRawText = token.Text.Replace(@"\n", "\" + Str.Lf + \"");
5      var updatedValue = token.ValueText.Replace("\n", "\" + Str.Lf + \"");
6      var newToken = SyntaxFactory.Literal(token.LeadingTrivia, updatedRawText, updatedValue,
   ↪ identifierToken.TrailingTrivia);
7
8      var sourceText = await expressionSyntax.SyntaxTree.GetTextAsync(cancellationToken);
9      return document.WithText(sourceText.WithChanges(new TextChange(token.FullSpan,
   ↪ newToken.ToFullString())));
10 }
```

Listing 3.9: Implementation of a Roslyn code fix to replace unescaped newline sequences in string literals

### 3.5.3 Using custom analyzers in RoslynPad

The project template for analyzers scaffolds two additional projects: one for a VSIX (Visual Studio integration extension) package, and one for a NuGet package. A VSIX package can be installed as a global extension in Visual Studio so that the analyzer is used for any kind of project. A NuGet package can be installed into specific .NET projects so that it applies only to the code in that project. Neither of those options are viable for use in LCDPs. A citizen developer doesn't work with Visual Studio, and the scripts they develop are not part of a .NET project.

While additional NuGet packages can be loaded by RoslynPad, the feature is not available in the editor component used in this example. RoslynPad has also rudimentary support for the built-in Roslyn analyzers [6], but there is no official way to add custom analyzers to the editor. However, as the analyzers and code fixes itself are just regular assemblies, they can be added to Roslyn instances independently.

This can be achieved in the same way RoslynPad adds the existing Roslyn analyzers internally, and requires only several lines of code that have to be adopted from its source code [7]. Listing 3.10 shows how the assemblies are added. The method `InitializeAnalyzers()` takes an instance of `IRoslynHost` and a `DocumentId`, both of which are created previously as part of the initialization of the editor. For each analyzer and code fix assembly, an instance of `AnalyzerFileReference` is created and added to Roslyn's internal project representation. As most of Roslyn's object structure is immutable, an updated instance of the `Project` object is returned and has to be integrated back into the original `RoslynHost` (lines 7-10).

```csharp
public void InitializeAnalyzers(IRoslynHost host, DocumentId documentId)
{
    var loader = host.GetService<IAnalyzerAssemblyLoader>();
    var analyzerReference = new AnalyzerFileReference("SampleAnalyzer.dll", loader);

    var document = host.GetDocument(documentId);
    var project = document.Project.AddAnalyzerReferences(new[] { analyzerReference });
    document = project.GetDocument(documentId);

    host.UpdateDocument(document);
}
```

Listing 3.10: Adding an analyzer to the RoslynPad editor

Correctly adding the analyzer assemblies enables RoslynPad to display them in the same way the default Roslyn analyzers are. Figure 3.13 shows how the analyzer is displayed and how a code fix is invoked on the code emitting the diagnostic warning. The custom code analyzer developed in section 3.5.1 is shown in figure 3.13a, the unescaped \n sequence is detected and underlined, with a tooltip showing details of the warning. Since a code fix is available, a yellow dot in the margin right of the line number indicates that the user can take an action to remedy the warning. Upon clicking the indicator, a context menu, shown in figure 3.13b, opens and shows the available actions. The first action, *Replace with compatible newlines*, is the code fix developed in section 3.5.2. The other three entries are refactoring actions coming with Roslyn by default. After the menu entry is selected, the code is changed according to the implementation of the code fix (figure 3.13c).

## 3.6 Summary and discussion

Chapter 3 introduced tools that aid citizen developers in writing code by extending the editor they are working with, and making it easier to understand APIs by wrapping them in more refined and approachable functions and providing additional help and documentation for frequently made mistakes. The augmentation framework developed in section 3.3 can be used to create customizable visual clues based on regular expressions in the AvalonEdit code editor. Using RoslynPad, an editor that is based on AvalonEdit and extends it with features of the Roslyn compiler platform, the integration of external API assemblies, custom static code analyzers and code fixes was demonstrated.

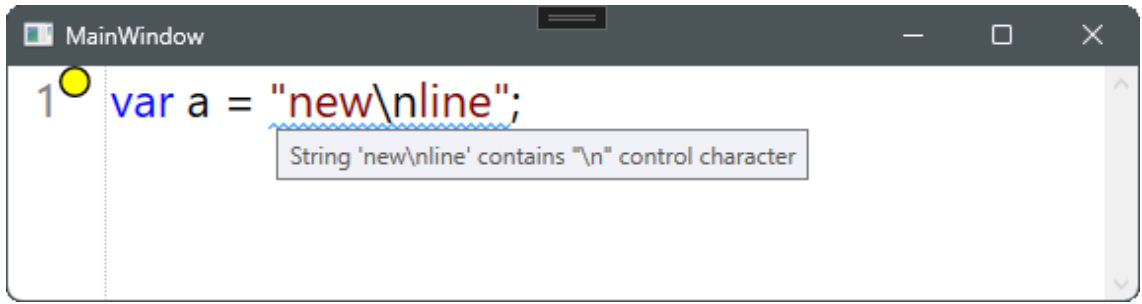

(a) Tooltip and indicator of the analyzer violation

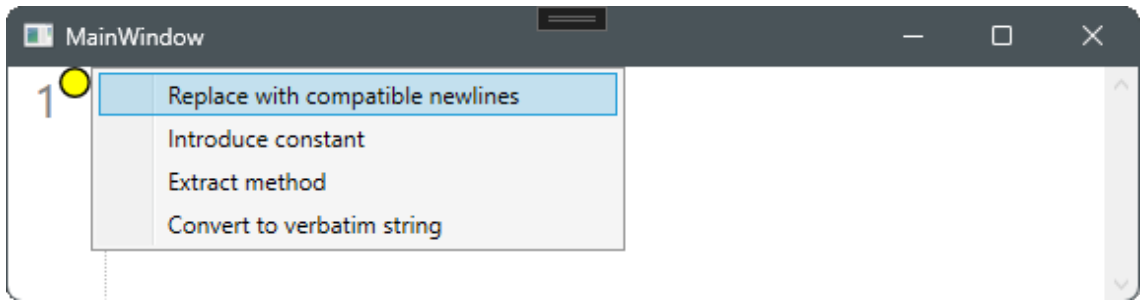

(b) Quick action menu showing possible code fix actions

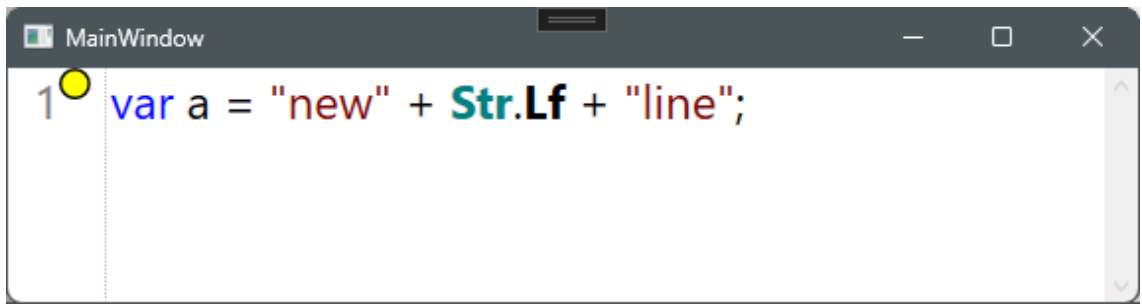

(c) Code after invoking the code fix

Figure 3.13: Custom analyzer and code fix in RoslynPad

These tools contribute primarily to the answer to RQ I posed in section 1.4. The augmentation framework provides an easy way for LCDP developers to build visual aids tailored to their users. Since the framework is specifically written for AvalonEdit's extensibility pipeline, its usefulness primarily depends on whether AvalonEdit is a viable editor component for the LCDP. Furthermore, since the framework uses regular expressions to determine where augmentations are displayed, more complex syntactic scenarios, such as detecting whether a code expression is inside a static method or part of an internal type, are difficult to detect. For this purpose, static code analysis, for example using the APIs exposed by Roslyn, is more suited. Building static code analysis rules and code fixes based on them is a popular function in many more advanced IDEs, but usually not present in simple code editor components for integration in existing applications. By using the existing scaffolding provided by Roslyn, an approach to develop and integrate analyzers and fixes into RoslynPad was demonstrated, which can be used to provide citizen developers with additional support on a syntactic level, albeit with very limited visual customizability. Future work could leverage static code analysis to build augmentations for more advanced syntactic structures.

Additionally, both the augmentation framework and wrapper APIs can be used to answer RQ II. By focusing on the detection of potentially insecure or harmful usage of APIs, users can be educated with appropriate messages while also offering them more secure alternatives. Specially prepared APIs, that wrap complex classes or functions susceptible to misuse, can contribute to security as well by providing users with readily made and easy to use abstractions. This, however, may require significant effort to develop and maintain. Unless the affected APIs can be clearly identified, each potentially insecure class or method must be considered for wrapping or has to emit warnings. New releases of frameworks must be analysed for new, previously not covered APIs. Furthermore, programming language features such as reflection or dynamic types have to be factored in, since they can be used to circumvent restricted APIs by loading or executing arbitrary code and expressions, both at compile- and runtime. Additional discussion of RQ II can be found in the summary of chapter 4 in section 4.6.

# Chapter 4

# Code security in low-code applications

Security flaws in LCDPs can potentially affect thousands [122] of applications, as vulnerabilities will be distributed with every instance of a LCA generated. This chapter will present two sources of possible security issues and examine potential mitigation strategies. The first source stems from the code embedded by citizen developers, as it can introduce unintended security problems, and should therefore be analysed for possible weaknesses during development time.

The other source of problems can originate on a higher level, when running a shared LCDP on a server for multiple, possibly independent, clients. To prevent users from accidentally or maliciously accessing system resources or data of other users, isolation of the different LCDP instances has to be ensured. This is even more important if the applications use sensitive company data, as data leaks would be critical and customers may demand an audit to prove that the environment is secure.

## 4.1 Process isolation

Several issues listed in the *OWASP Top 10 of Low-Code/No-Code Security Risks* [8] directly relate to the problems examined in this thesis, namely '*Data and Secret Handling Failures*', '*Security Misconfiguration*', '*Injection Handling Failures*' and '*Vulnerable and Untrusted Components*'. The former two risks are connected to the separation of LCDP instances on a server, while the latter two focus on issues with code written for the LCA. This section will explain the concepts of process isolation, to separate running applications from each other and prevent access to foreign data.

Process isolation can be implemented on hard- and software-level. Aiken et al. [2] define it for the general case as:

> ' *Isolation protects system integrity by preventing one process from interfering with another's, or the system's, code or data, and by preventing untrusted code from accessing protected resources. Isolation also contributes to system resilience by providing failure boundaries that permit part of a system to fail without compromising the whole.* '

On hardware level, this isolation is enforced by the *memory management unit* (MMU) by preventing a process from accessing memory pages outside those allocated to him [148, p. 260] as well as running user processes at a lower permission level than kernel processes (which have the most privileges) [148, p. 479]. High-level programming languages (such as C#) usually do not have access to these hardware intrinsics (outside of device drivers), since they run in *user mode*, where direct access to the operating system kernel and underlying hardware is not possible [108]. However, this only protects processes and the kernel on a relatively low level and does not stop an application from accessing foreign data that resides, for example, on a hard drive or on a network server. On operating-system level, further access control models, such as *discretionary access control* (DAC) [148, p. 612] or mandatory access control (MAC) [148, p. 612] may be implemented to selectively allow or deny users access to files and resources or enforce a flow of information. While these may prevent accidental access to data, they do not provide isolation from failure. As Tanenbaum points out that '*All these ideas about formal models and provably secure systems sound great, but do they actually work? In a word: No.*' [148, p. 615], a more thorough concept for both security and failure is needed.

LCAs, with a mixture of generated and user-written code, can be seen as *mobile code*: code from an external source that is executed as part of an application the user started [65] [148, p. 698]. For example, suppose a user is running a LCA that has been developed collaboratively with colleagues. The user is reasonably sure that none of the colleagues had deliberately included malicious code, but is uncertain whether the application makes use of insecure external code that may have been loaded by a third-party library. Bargury et al. [8] describe this risk as *dependency injection*. Wahbe et al. [153] highlight a similar issue, where (poorly implemented) additional software modules purchased for an application can make the main application seem unreliable. They propose *sandboxing* as a solution, where an untrusted module is encapsulated by adding additional instructions so that it can only operate within its address space and not accidentally alter trusted code. A more modern definition from Hoopes [59] describes a sandbox as '*a monitored and controlled environment, such that the unknown software cannot do any harm to the real hosting computer system*'.

Sandboxing is common today, for example in browsers such as Chromium. To prevent malicious code from affecting the user's data, security-critical components are run in the sandboxed *rendering engine*, which is the common target for exploiting browsers vulnerabilities [55]. The rendering engine runs in a separate *target* process, while the main browser process is called the *broker* and manages the sandboxed processes [51]. Figure 4.1 shows the data flow of this architecture: the broker sends the source code of a web page (containing HTML and JavaScript) via inter-process communication (IPC) to the renderer, which in turn generates the web document and sends it back to the browser process for displaying [9].

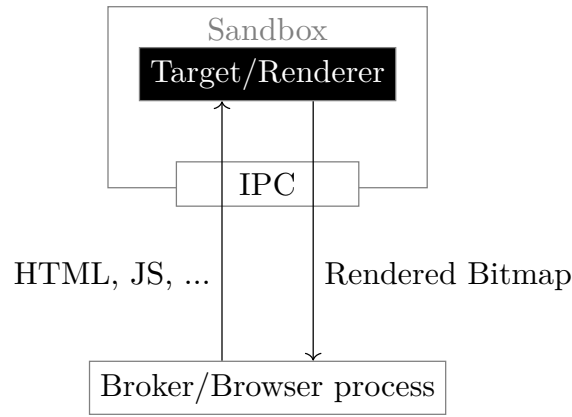

Figure 4.1: Sandboxing architecture in Chromium. Adapted from [9] with terminology from [51].

Barth et al. [9] note that this architecture does not prevent attacks on the renderer by itself, but hinders an attacker from accessing the user's data outside the sandbox. Chromium's sandbox contains the malicious code, prevents execution of potentially damaging system calls, and limits it to using the browser process's interface, which provides a smaller attack surface. More generally, a sandbox itself doesn't prevent malicious code from executing, but tries to reduce the damage it can inflict by intercepting and mediating system calls (such as direct access to the file system, or creating network connections) [55]. It does not prevent attacks on components outside the process. An example for this are the CVE (Common Vulnerabilities and Exposures)[1] issues CVE-2010-2897 [30] and CVE-2010-2898 [31], which use specially prepared CSS to target an issue in the Windows kernel's font handling, and JavaScript code to cause a crash in glibc[2] with very long file paths respectively.

## 4.2 Threat model

Running a shared LCDP on a server for multiple, independent clients is similar to a Platform-as-a-Service (PaaS) offering in a cloud environment. Depending on the implementation, clouds may share the software resources, such as programming libraries and database engines, between users, or provide isolated environments using virtual machines (VMs) for each tenant [132]. Using virtual machines to isolate individual instances is considered the primary defence mechanism against attackers in the cloud [44], but moves the responsibility for security to the virtualization layer. This presents different threats, for example cross-VM attacks using side-channels to gain information on other VMs running on the same system. These can be used to steal cryptographic keys, inhibit the performance of the attacked VMs or perform denial-of-service attacks [131].

---

[1] A system to catalogue security vulnerabilities — https://cve.org
[2] The C standard library implementation for GNU/Linux.

Since the primary concern of this thesis is the code security of LCDPs, a threat model that focuses on risks originating from additional source code written for LCAs is introduced. Whether the application is hosted in a private or public cloud, on premises or in an off-site data centre, an obvious threat is malicious behaviour by either the provider of the services itself, or single actors involved in the administration of the servers. For the purposes of this model, the infrastructure, including possibly used virtual machines, and involved parties are considered trusted. The LCDP is hosted on a single server, set up by a vendor, and offered to customers. Customers use the LCDP to develop and test generated applications (introduced in section 2.2.3). The resulting applications are however not hosted by the vendor, the customers deploy their LCAs individually as part of their infrastructure. This limits the attack surface to the development time of the LCAs, and the possible attackers to those individuals who use the LCDP to built applications. While a generated application may contain code that targets the customer after it is deployed on their infrastructure, an attacker would ultimately try to compromise their own data. In this model, the customer trusts their developers to not threaten their own data.

The main interest of the LCDP vendor is thus to isolate each instance of the platform against each other. A naïve attack can be, for example, to try to access files and folders outside the directories required for development. Suppose the LCDP is running on a web server on a Unix system. Customers each have their own folder on the server: `/var/www/customer-one`, `/var/www/customer-two`, `/var/www/customer-three` and so on. Unless the server administrator has set appropriate protection bits [148, p. 45] on the folders, so only developers of the respective customer can access a directory, each customer could peek into the files of the others. Current web servers usually provide easy means to set up applications in isolated directories to prevent such issues, IIS[3] for example, separates individual websites into *application pools* [91].

Another possible attack vector is a denial-of-service attack. An adversary could try to exhaust the system resources of the server by spawning numerous processes with the aim of preventing other customers to be able to use their application instances on the server. This scenario can also occur accidentally, when malfunctioning code contains a memory leak and continuously allocates memory until all available RAM on the system is in use.

---

[3] Internet Information Services, a web server by Microsoft

## 4.3 Approaches for securely running code

There are a few general methods for the isolation of code running as part of processes and limiting the impact that code can have on its environment. Neither of these approaches are exclusive to low-code, as in the end, LCAs on a fundamental level are not different from other applications. It is important to distinguish between the two kinds of code that comprise a LCA: the code that is generated by the LCDP (or runs as part of its environment), and the custom code written by the user. Some security technologies, such as sandboxes, often require usage of specific APIs to profit from it [55]. For the code that is part of the LCDP, this can be feasible, if it has already been built with specific security platforms in mind. Otherwise, it would require substantial changes to existing code. Implementing *AppContainers*, which are described in detail in section 4.5.3, requires usage of security tokens and running additional processes or threads, which are no small architectural changes.

And since the citizen developer cannot be expected to write specifically secure code, or programming against an API facilitating security, relying on securely written code should not be the primary guard against accidental or malicious threats. Viewing the entire application as untrusted, and isolating it from the operating system and other processes, can help to prevent undesired access to system resources [55]. This section investigates some commonly available techniques to run applications in an isolated environment.

### 4.3.1 Sandboxing

Sandboxing can refer to a broad number of techniques that transparently verify or control the data flowing to and from an application, with the aim to prevent interference with other parts of the system. According to Goldberg et al. [50], who '*use the term sandboxing to describe the concept of confining a helper application to a restricted environment, within which it has free reign*', the name originates from Wahbe et al. [153], who used it in the context of sanitizing memory addresses. Al Ameiri et al. [4] distinguish between six types of sandboxes:

**Applets** are run inside an environment that is responsible for isolation, such as Java Applets or Flash objects in a browser.

**Jails** verify and potentially block system calls and decide about access to resources of a '*prisoner*' process [148, p. 695].

**Virtual machines** provide an entire guest operating system for an application, and enables access to system resources through a virtualization layer.

**Rule-based execution** lets a user specify how an application is allowed to interact with the system.

**Standalone applications** provide an additional layer above the operating system that intercepts activity of an application and redirects it into isolated objects, so it can't affect the host system. Examples are Sandboxie [136] for Windows or Firejail for Linux [118].

**OS-integrated** tools to support sandboxing, such as *AASandbox* for Android [15], which performs static and dynamic program analysis to detect potential malicious use of system calls, or *seccomp* [71], a Linux kernel module, which enables applications to filter incoming system calls.

It should be noted that their evaluation preceded the broad use of containerization tools like Docker, which was first released in 2013, that utilize operating system technologies to partition a host OS to share its resources between multiple containers, and thus would fall into the last category.

There is a certain overlap between these categories. For example, jails and operating-system integrated tools both control access to system calls by the sandboxed application. Rule-based execution modules, such as SELinux and AppArmor [4] for Linux systems and AppContainers under Windows, all utilize mandatory access control and are closely integrated with the operating system as well. Java Applets and Silverlight applications run inside the Java Virtual Machine (JVM) and Common Language Runtime (CLR) respectively. Both are examples of high-level language virtual machines (which in turn are a kind of process VM) [144, p. 228], while VMs running a guest operating system on a host are called system VMs.

In the context of this thesis, some of these categories can be eliminated as a potential choice to securely run LCAs. *Applets* are highly dependent on the used programming language and target platform, and require an environment that supervises its applications to be supplied. Specifically, the mentioned technologies Java Applets [123] and Flash [1] have both been deprecated for several years and were mainly used in web browsers. *Jails* originate from Unix- and BSD-like operating systems and are not universally available, but can be categorized as an early implementation of containers [28]. Tools integrated into the operating system facilitate sandboxes, but, in the case of *seccomp* for example, aren't complete sandboxes: '*System call filtering isn't a sandbox.*[...] *It is meant to be a tool for sandbox developers to use.*' [71].

Virtual machines and standalone applications are interesting options, since they don't need to be specifically supported by the application to be sandboxed, are available on the most common operating systems and not limited to certain types of applications. The drawback of traditional VM solutions, like VMware, is the performance loss of about 10-20% [42] when compared to a native system. This approach is also relatively heavyweight, requiring an entire OS installation per application.

A compromise between ease of use and system requirements is OS-level virtualization. Regular virtualization often uses a hypervisor, or virtual machine monitor (depending on the hard- and software capabilities and configuration), to interface the virtual machines with the hardware of the computer, either directly on the hardware (type I) or on a host OS (type II). The VMs itself always contain a guest operating system. OS-level virtualization shares the resources of a host OS, including the kernel, among all virtualized instances (containers) [12]. The host OS provides the facilities to implement these containers, for example, Docker utilizes the seccomp kernel module [40] mentioned earlier. Figure 4.2 illustrates the difference between the two types of virtualization. Two concrete implementations of such containers are discussed in detail in section 4.4.

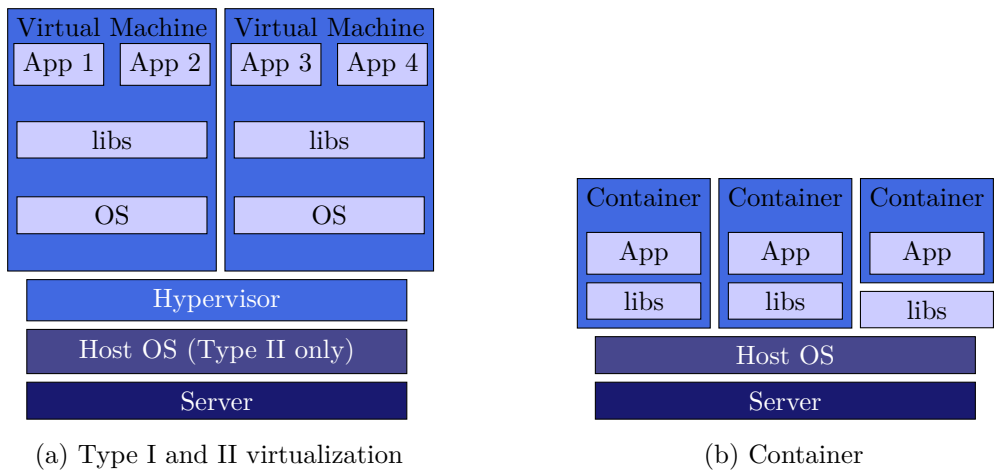

(a) Type I and II virtualization

(b) Container

Figure 4.2: Types of virtualization. Adapted from [12]

### 4.3.2 Blocking of API calls

To prevent users from calling certain APIs that can be used to potentially compromise a system, an obvious approach may be to simply disallow usage of the methods in question. For example, APIs that access the file system, security-relevant information or enable usage of reflection, could be detected and blocked from execution, or prevent compilation altogether.

Online services that provide compilation of code in a browser, such as SharpLab[4] and .NET Fiddle[5], which enable testing of small code fragments or analyse the resulting machine code, have a similar problem — the server their software is run on should not be able to be compromised by a malicious actor. Historically, SharpLab used the *Unbreakable* library[6] to selectively allow certain APIs to be called and imposed limits on execution time and memory usage. This approach was abandoned, as it didn't scale well and required continuous maintenance of the allowed or denied APIs [142]. Its current implementation uses AppContainers, which are detailed in section 4.5.3.

---

[4] https://sharplab.io
[5] https://dotnetfiddle.net
[6] https://github.com/ashmind/Unbreakable

Both maintaining a list of allowed APIs (*allow list*), denying everything by default, or a list of denied APIs (*deny list*), allowing everything else by default, has its problems. An allow list has to have an appropriate amount of default classes and methods, that are widely used for normal programming tasks. This would have to include a vast amount of the basic API provided by the framework, and can potentially frustrate users if safe classes and methods were forgotten or have to be manually enabled. Likewise, a deny list would have to contain a comprehensive amount of abusable APIs, including reflection, file input and output or the creation of processes. Classes for reading and writing files, for example, can be used both for valid business cases (such as creating files in an allowed folder) and malicious purposes (reading files outside the user's directory). On API level, it cannot be decided whether calls like this are benign or not. As such, additional security measures, like filesystem-level access control, have to be employed.

Future releases of an API or its framework have to be scrutinized for new method calls, or changes in existing behaviour that can be either beneficial to the user (and need to be allowed) or can be misused (and need to be denied). A single oversight can potentially allow the lists to be circumvented altogether.

If the intent is to steer the user away from using certain APIs (because they may be obsolete or insecure), an alternative approach is to use context-based advice, as introduced in section 3.4.2.

## 4.4  Implementations of OS-level virtualization

To test the suitability for the deployment of low-code applications, two implementations of sandboxing environments are evaluated: Windows Sandbox and Docker containers. The following sections explain their isolation and security architecture, effort required to customize and set up an instance, and their ability to run a LCA desktop and web application on demand. The two applications were tested on a system with an AMD Ryzen 5 3600 CPU, 32 GB RAM, Windows 11 Pro (Build 22623) and an internet connection with 50 Mbit downstream.

### 4.4.1  Windows Sandbox

The Windows Sandbox is a lightweight environment for the isolation of applications for Windows 10 and 11 that runs directly under Windows and uses hardware-based virtualization (Hyper-V) to run a separate kernel for each instance [113]. Instances of the sandbox are transient, i.e., after closing the sandbox its state is not persisted (except for reboots initiated from within the sandbox) and each new instance gets loaded with an unmodified installation of Windows. It is possible to map folders from the host system into the sandbox, and to execute commands upon loading to automatically install applications or set up a specific environment configuration. The sandbox has network access, and can use resources, such as audio and video input, printers, a battery if present, and even the GPU from the host system. [114, 115].

**Architecture**

Windows Sandbox shares part of the system files with its host system. Microsoft calls this a *dynamic base image* [114] and differentiates between mutable and immutable system files. Most of the files the OS uses are read-only (immutable), and thus can be shared safely with the sandbox. The files required to be modifiable (mutable) are copied to the sandbox, so that changes don't affect the host OS. The immutable files are only linked into the sandbox, using a multi-layered structure [61], that draws the files from several sources in the host system.

The immutable system files are also shared in memory, called a *direct map* [114], reusing the same pages to reduce the amount of memory the sandbox requires. The sandbox also doesn't use a static amount of memory from the host OS, and behaves more similar to a regular process. If the host needs to reclaim memory, it can reallocate memory from the sandbox to the processes that require it.

Using *integrated scheduling* [114], the process scheduler of the host can treat processes inside the sandbox as threads and prioritize them among the workload of the entire system. In traditional VMs, the hypervisor controls the processors assigned to the virtual machine in its entirety, and has no information or influence about the processes running inside it.

**Customization**

By default, there are a limited number of configuration options for the sandbox. Administrators can configure the mapped folders into the host system, the commands to be executed upon logon, disable redirection of certain resources and the amount of memory available to the sandbox [115].

While there was previous effort to customize the image that the sandbox uses as a base [61], the steps outlined in the article could not be reproduced on Windows 11, due to apparent changes in the file and folder structure of the base files. However, placing additional files in `%ProgramData%\Microsoft\Windows\Containers\BaseImages\<randomGUID>\BaseLayer\Files\` will make them available inside the sandbox. Tests with a simple text file and a portable version of Notepad++ showed that files and applications added to the sandbox this way are writable and executable, changes to them are contained and not propagated to the host system. While copying files or installing programs can be achieved by running a script on login as well, it would take more time, since the files are not immediately available as part of the base files. Since this is not an officially documented way to customize the sandbox, it is a volatile approach which may not work after subsequent updates to Windows.

**Usage**

Reiterating from section 4.2, the threat model involves adding malicious code to a generated LCA and trying to gain access to other data on the server, or disrupting other users working on the same server. Isolating the generated application involves deploying it to a new instance of the Windows Sandbox. Depending on the type of application, this can be relatively easily achieved by configuring a mapped folder to the host system, which contains the generated application, or using a script to copy and install the application and its dependencies.

For desktop applications that provide a graphical user interface (GUI), the sandbox would have to be started on a remote server, and the users would need a way to connect to it to test their application. A naïve approach, using an RDP (Remote Desktop Protocol) connection and the built-in Windows RDP functionality, was unsuccessful, as no connection could be established, although the sandbox was reachable in the network and had remote access enabled. A further test with the third-party tool TeamViewer[7] succeeded. This, however, involves additional (paid) software which has also to be deployed in the sandbox and on the computer of the client, and adds more complexity and reduces user experience.

A test with a simple ASP .NET Core web application that runs on the integrated, in-process Kestrel web server in the sandbox, was more successful. If configured to accept external connections, it was possible to access the hosted website from different computers in the same network.

**Evaluation**

Whether Windows Sandbox is a suitable environment to deploy LCAs into largely depends on the impact of the temporary nature of the environment. Since the Windows installation inside the sandbox has to be configured on every boot and applications have to be deployed each time, setting up the sandbox can potentially take a long time, if many or large programs have to be installed. Windows Sandbox does not officially support snapshots or states to start from; thus, a once configured system cannot be reused. Starting the sandbox itself is relatively quick, requiring about six seconds until the desktop is visible and usable. However, only a single instance can be run concurrently — trying to open a second instance fails with the error message, that only one sandbox can be active at a time.

Mapped folders of the host can present a potential security risk, if they contain sensitive data, which could be read by an untrusted application in the sandbox, or are writable, allowing an untrusted application to break into the host system. Network access is by default available in the sandbox, enabling threats to the local network or downloading additional malicious code from the internet. As the memory used by the sandbox can be capped via configuration, memory exhaustion attacks can be mitigated. Unless limited by external means, programs running in the sandbox can fully utilize the CPU of the host machine and potentially affect other applications running on the system.

---

[7] A client-server based remote control software — https://www.teamviewer.com

Deploying and using applications that require a GUI is possible, but requires an external tool for users to be able to connect to the sandbox and interact with the generated application. Hosting web applications inside the sandbox is easier and requires only minimal configuration effort. However, it is questionable if running sandboxes on more than a small scale is feasible — instances have to be manually started and configured, and there are no tools supplied to facilitate automated deployment and configuration management. Windows Sandbox is not advertised as a commercial-grade OS-level virtualization solution, and the focus on easy usage and minimal configuration options, as well as the absence of snapshotting and provisioning options, makes it generally more suitable for individual users who require the execution of untrusted applications on a local computer with minimal effort.

### 4.4.2 Docker

Docker is a software suite for the creation, management, and execution of containers [28]. Containers allow for a deterministic and standardized way to define environments in which applications and their dependencies are installed, and can be used to distribute and deploy them on local or remote, physical or virtual machines [34]. Docker builds upon multiple security features that are available in the Linux kernel: *cgroups*[8], *namespaces* and seccomp [147]. Cgroups are used to limit and isolate the availability of system resources for a collection of processes [77], namespaces enable the assignment of resources to processes without affecting their usage by other processes outside the namespace [72], and seccomp blocks the usage of certain system calls.

**Architecture**

To create, administer and run containers, Docker uses a client-server architecture consisting of the Docker client (`docker`), a daemon (`dockerd`), and a *registry*. The Docker daemon runs on the host on which the containers will be instantiated and provides a REST API (UNIX sockets or network interfaces are also possible) to enable communication with the client. The registry stores *images* and makes them available to *pull* from the host machines. Registries can be publicly available or run privately. Images are templates that define how a container will be created, and contain information such as the underlying operating system, commands to install additional software and further configuration, like the assignment of environment variables. They can be hierarchically connected, so that commonly used operating systems or software are available as a basis for the creation of user-specific images [34]. Official images for operating systems like Windows or various Linux distributions like Debian or Fedora are distributed by their respective developers, and additional images with preconfigured software environments are often available as well. For example, the official Microsoft Artifact Registry[9] supplies images for various Windows and Windows Server systems, .NET SDKs and other tools that can be combined as needed.

---
[8] Short for *Control Groups* [77]

Figure 4.3 shows a high-level overview of Docker and the relationship of basic commands with Docker's components. The following two sections describe how containers can be customized and used for the deployment of LCDPs and LCAs.

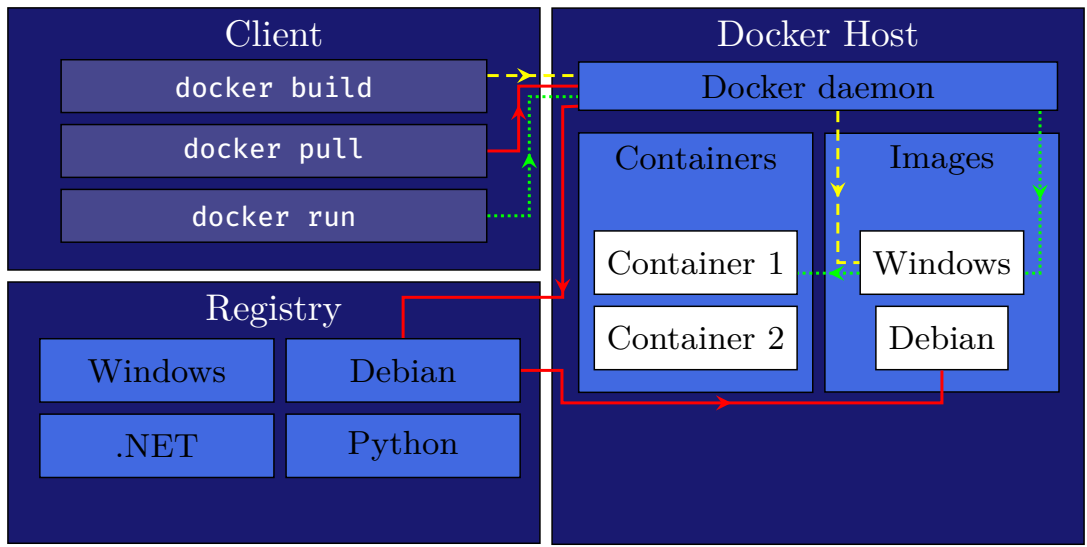

Figure 4.3: Architecture of Docker. Adapted from [34]

While Docker utilizes Linux kernel features to implement OS-level virtualization, it also runs on Windows and macOS. Historically, Docker used Hyper-V to run a small Linux VM to provide the ability to run Linux containers on Windows. After Microsoft introduced the second version of their Windows Subsystem for Linux (WSL2)[10], Docker switched its integration on Windows to use WSL2 instead of their own VM [45]. Conceptually, WSL2 still uses Hyper-V to run a Linux in a lightweight VM internally [92], however since it is much more integrated with Windows, it provides better performance and more capabilities than Docker's original implementation [45].

**Customization**

Docker images are defined by templates called *Dockerfiles* that consist of commands which specify how a container is assembled [37]. The `docker build` command, as visualized in figure 4.3, instructs the Docker daemon to build an image from a Dockerfile. Listing 4.1 shows a sample Dockerfile which compiles and deploys an ASP.NET Core web application into a container. The Dockerfile consists of two different stages, which are defined by the `FROM ... AS` statements: one named `build`, and another unnamed stage (that can be implicitly referred to by an integer, starting with 0). Each statement in the file adds another *layer* to the image. Since Docker caches images internally in a *build cache*, unchanged layers are reused on subsequent `docker build` calls, reducing the time required to update and rebuild images [36].

---

[9] https://mcr.microsoft.com
[10] A feature to run a Linux system inside Windows

```dockerfile
# Set base image for building project
FROM mcr.microsoft.com/dotnet/sdk:7.0 AS build
WORKDIR /source

# Copy project files and restore packages
COPY exampleProject/*.csproj .
RUN dotnet restore --use-current-runtime

# Copy remaining source of application and publish it
COPY exampleProject/. .
RUN dotnet publish -c Release -o /outputFolder --use-current-runtime --self-contained false
    --no-restore

# Set base image for runtime project and copy published application from build stage
FROM mcr.microsoft.com/dotnet/aspnet:7.0
WORKDIR /outputFolder
COPY --from=build /outputFolder .
ENTRYPOINT ["dotnet", "exampleProject.dll"]
```

Listing 4.1: Dockerfile for compiling and deploying an ASP.NET Core application. Modified from [38]

The `FROM` statement defines what image will be the *base image* for the container. Since the `build` stage compiles the ASP.NET Core project, it will be based on the most recently available .NET SDK version (7.0) image. In the next step, the project files (`*.csproj`) are copied to the image, and the packages containing dependencies for all projects are restored. As the project files usually only change when dependencies are added or updated, the image only changes infrequently up to this point and is thus cached and reused by Docker. After the dependencies are restored, the remaining source files are copied to the image and the application is compiled and published to an output folder. Running the application does not require the full .NET SDK; the second, unnamed stage is thus based on the `aspnet:7.0` image, which contains only the required the ASP.NET Core and .NET runtime libraries. The published application is then copied from the previous `build` stage into the final image. The `ENTRYPOINT` statement defines a default command that is executed when the container is run, in this case `dotnet exampleProject.dll`, which starts the application on a local web server.

Dockerfiles can run arbitrary programs and scripts using the `RUN` statement, which enables to set up containers with any complexity. However, it is recommended to keep the containers as stateless and '*ephemeral*' [35] as possible, since they should be able to be started, customized and stopped with minimal configuration effort. Data that needs to be saved persistently should be stored by a dedicated, separate service, such as a database, which in turn can be stored on the host's filesystem using *volumes* [41].

By default, Docker puts no constraints on the usage of resources by containers. If a container begins to exhaust system resources, the underlying operating system can act to ensure that no process compromises the availability of the system. For example, the Linux kernel will start to kill processes based on individual runtime and memory usage when not enough memory is available [52, p. 194]. As this may cause the termination of any process (potentially a non-container process) on the system, Docker allows to put limits on memory-, CPU-, and GPU-usage per container [39].

**Usage**

Based on the threat model from section 4.2, a generated LCA should not be able to interfere with other users or data on the server during its development time. Deployment of the generated application to the container works similar to the process described in the previous section. A generic Dockerfile using a suitable base image with commands to build the LCA from source and deploy it into the container is created once and can be used for all customers of the LCDP, provided the generation of the LCA, including all dependencies, is standardized across the platform. Since Dockerfiles can be parameterized, variable values like file or folder names can be set as placeholders and passed when building the image [37]. If customers can extend their LCA with custom dependencies or require special set up steps, the Dockerfile could be expanded with additional scripts, that are executed at fixed steps during the creation of the container (for example *before publish* or *after copying build artefacts*) and act as extension points.

After the image is built by the `docker build` command, it can be shared among multiple clients. Images can either be exported and imported manually, or stored in a public or private registry. A client can then pull and run images, as shown in figure 4.3. `docker pull` instructs the daemon to find a specific image in a registry and download it to the host. Alternatively, `docker run` can be used directly to instantiate a container, whose image is automatically pulled from the registry as needed.

During the active development of a LCA, frequent code changes and rebuilds, as with traditionally developed applications, are to be expected. This means that containers are relatively short-lived and often stopped, reconfigured and restarted. Depending on the scale of the LCDP and the number of concurrent users, a single server may not be enough to run containers from all users at the same time. Since one container usually only runs one process, and an application may consist of several processes (such as a database server and a web server) [116], redeploying a LCA can become more complex, as multiple containers have to be coordinated.

Management of multiple containers at scale is the task of *container orchestration engines*, such as Docker Swarm[11] and Kubernetes[12]. Common features are scheduling, grouping and moving containers between different machines, health- and integrity monitoring, and guaranteeing their availability [116].

**Evaluation**

Architecturally, Docker is much more suited to quickly deploy and test generated LCA on-demand than the previously discussed Windows Sandbox. The ability to define and build images that include the required dependencies for generated applications provides a deterministic way to run a stable system environment. The availability of different operating systems as base images, and images of various programming frameworks and runtime environments that build upon them, make it easy to offer the user a variety of platforms to deploy their application on. As the images are built with layers, which also support caching, recompiling only the changed components of a LCA enables quick recreation of a container after changes are made to an application.

To get an estimate of the amount of time required, the basic commands required to set up a Docker container were benchmarked. The components of the testing system are described in section 4.4. Each command was run three times to determine the median and average. The example Dockerfile from listing 4.1 with a minimal ASP.NET Core web application was used to perform the tests. The timing was determined by the built-in measurements each docker commands prints after it has finished. The results are shown in table 4.1.

The time required to build and run a container largely depends on two factors: the size of the base image, and the complexity of the application to build. As once downloaded base images and unchanged layers are cached, subsequent builds of a container are much faster. While the building of the initial image took around one minute, updates to the image were finished in a reasonable 7 seconds. Running the container was practically instantaneous, the integrated web server run by the `dotnet` command was started and available almost immediately after startup of the container. More complex applications, that require longer compile or build times, will, of course, result in longer container deployment times, but the overhead incurred by docker in this simple example was within reasonable bounds.

Whether Docker is an appropriate approach to run generated LCAs depends mostly on the kind of application. By default, most container images only provide a command line interface to interact with the system running in a container instance. This means that applications requiring interaction via a user interface are generally not directly usable, as there is no *desktop* environment available that a client can connect to. Applications that are primarily server-based (such as an ASP.NET Core application running on a web server, similar to the environment defined in listing 4.1), and can be accessed, for example, via browser or network, are better suited to run on this infrastructure.

---

[11] https://docs.docker.com/engine/swarm
[12] https://kubernetes.io

| Command | Description | Median/Average |
|---|---|---|
| `docker builder prune -f`<br>`docker build --pull --no-cache .` | Build a docker image without using caches or existing base images | 58.7s/60.3s |
| `docker build .` | Build a docker image with trivial source code change | 7s/7.1s |
| `docker run thesis/bench:1.0` | Start the previously built image named `thesis/bench:1.0` | No measurement provided by docker, but less than a second. |

Table 4.1: Benchmark of basic Docker commands

There are however tools available to make GUI access available for Linux containers, for example *x11docker*[13], which involves starting a display server inside the container and then connecting via a window client or other remote control software. This is not possible with Windows containers. However, Microsoft provides a Windows Server image with *desktop experience*, which makes UI APIs available inside the container. While there is still no GUI to interact with, it enables applications to utilize the GPU and consume the newly provided APIs [5].

Similarly to virtual machines, Docker containers need a dedicated system to run on, as users of the LCDP cannot be expected to host containers on their own machines. Containers are more lightweight than VMs, as downloaded base images and layers can be shared among multiple instances of a container. Since each container shares the kernel of the host OS, a compromised host kernel can potentially affect all containers running on a system. Virtual machines provide more isolation in that scenario, as the entire OS and kernel is contained inside the VM. However, running a VM for only a single application is inefficient and requires significantly more resources [74]. If the reduced isolation and lack of options to run GUI applications in containers is an acceptable tradeoff compared to the large overhead of VMs, containers are the more optimal solution when hosting a LCDP and building generated LCA for its users.

---

[13] https://github.com/mviereck/x11docker

## 4.5 Securely running code in C#

The .NET Framework, .NET Core and .NET contain several APIs that aim to provide security policies for and isolation between applications. Most of these technologies were used historically to manage access and permissions of assemblies, or separation of processes in memory. Microsoft discourages use of these APIs and recommends to use security features on operating system level, such as containers or virtualization [91]. This section gives a short overview over these deprecated APIs, and introduces a modern alternative that can be used instead.

### 4.5.1 Code Access Security

Code Access Security (CAS) is an, from .NET Core onwards obsolete [48], API that controls execution of code based on *evidence* and *permissions* [76]. Evidence contains information about the origin of the code, such as a website, directory, publisher, or a hash code. Permissions define what resources the code is allowed to access, such as printers, file input/output, or the Windows registry. To match evidence to permissions, *code groups* can be created, so that, for example, applications originating from a web source are disallowed access to local files by default.

CAS was deemed as too complicated to configure correctly by Microsoft [101], and since it is also limited to managed code, it does not affect native applications running outside the CLR, making the security guarantees it aims to provide flawed.

### 4.5.2 Application domains

Application domains (usually referred to as *AppDomains*) are logical containers that can be used to run multiple applications inside a single process [85]. Domains can be used to load assemblies, either *neutral*, for all domains with the same security parameters, or only for the current domain. Applications inside a domain are isolated from the other domains in a process, they cannot directly communicate, and faults in one domain do not affect others. Individual domains can also be loaded and unloaded at runtime, which is useful for isolating untrusted code in a domain and only allowing it to communicate with other domains via common inter-process mechanisms, such as pipes or remote procedure calls (RPC). Support for application domains was dropped from .NET Core onwards for being resource intensive [84] and not providing adequate security boundaries [117].

### 4.5.3 AppContainer

Compared to CAS and application domains, which are limited to applications running with the CLR, an AppContainer provides security and isolation on operating system level and can be used with existing and new applications. The Windows version of Google's Chromium browser uses AppContainers as part of its security infrastructure [51].

This section provides an overview on how AppContainers work, explains the basic steps necessary to run an arbitrary application inside an AppContainer using C# and evaluates its usage as a viable option to isolate applications. The full source code of the example is also available on GitHub [17].

**Background**

Current versions of Windows use mandatory integrity control (MIC) [96], an implementation of mandatory access control for processes, to define an integrity level (IL) for objects, which represent their trustworthiness [111]. *Objects* in this context generally refer to system resources, such as files, processes, or devices [94]. Four integrity levels are used: low, medium, high, and system. System is reserved for system processes, high integrity is assigned to administrators and elevated users, medium is the default level. Low-integrity processes are thus heavily constrained in their functionality, since they cannot access objects above their integrity level, unless they are explicitly granted rights to do so.

Windows uses secure identifiers (SIDs) to identify users and groups as trustees for an object [102]. SIDs are generated when a user or group is created and consists of several components, the most important being the value for a subauthority, which identifies a user in a domain and is represented by a globally unique identifier (GUID). A complete SID might look as follows: `S-1-5-21-4159745769-645805219-203622070-1001`. In addition to these generated SIDs, a number of constant *well-known* SIDs exist, which are used to identify special entities, such as a group that contains all users, or all administrators in a domain.

An AppContainer is assigned a SID as well, however it is not randomly generated, but calculated by a cryptographic hash function based on its name [26]. As this enables the SID to be persistent and reproducible, the identity of an AppContainer can be used to explicitly put applications into a specific container.

What a process inside a container is allowed to do is defined by a set of *capabilities*, that are also represented by well-known SIDs[14]. These capabilities are similar to the permissions assigned to apps in operating systems for smartphones, such as access to the internet or the music or picture library. More Windows-specific capabilities are the permission to access removable devices, or usage of user credentials. Access to specific files and folders can be granted by adding the SID of the container to the access control list (ACL) of an object.

---

[14] A full list of available capabilities can be derived from [109], but as there is no conclusive documentation, the author assumes that the enum values 84 through 94 apply to AppContainers, as they are explicitly mentioned for these SIDs

**Implementation of an AppContainer**

AppContainers were originally used for Windows Store applications that were introduced in Windows 8 and then known as *Metro* apps [107], or more recently as Universal Windows Platform (UWP) apps. These apps are automatically run in an AppContainer and don't have to specifically implement it. As there exist no managed APIs to create or manipulate AppContainers, the majority of the code requires usage of *Platform Invoke* (P/Invoke) [97], to call functions from the unmanaged Windows-API.

Since there is very little documentation on how to use AppContainers for non-UWP executables (besides a rough outline of the concept [95]), the correct usage of the unmanaged API of the following sample implementation is partially based on an existing demonstration project [27]. The sample implementation uses the Vanara[15] library, a thin wrapper around the native API, to simplify the otherwise verbose object declarations without changing their original names. The following samples omit error handling to improve readability.

```csharp
private AdvApi32.SafeAllocatedSID CreateOrGetAppContainerProfile(string containerName, string
    containerDisplayName, string containerDescription)
{
    var result = UserEnv.CreateAppContainerProfile(containerName, containerDisplayName,
    containerDescription, null, 0, out var sid);

    if (result == HRESULT.S_OK) return sid;

    UserEnv.DeriveAppContainerSidFromAppContainerName(containerName, out var existingSid);
    return existingSid;
}
```

Listing 4.2: Creation or retrieval of an AppContainer SID

The first step, shown in listing 4.2, is to retrieve the SID of the AppContainer to use. Depending on whether the container already exists, either a new container is created, or the SID of the existing container is derived from its name. The result of the method is a handle to the SID of the container. The next step is to build an instance of the `STARTUPINFOEX` structure, which is later used to pass initialization data, including information about the AppContainer, to the process that will run in the container. The capabilities are stored in a `SECURITY_CAPABILITIES` structure, which has to be populated before being used later in the process creation. As shown in listing 4.3, the SID of the container gets assigned to the `SECURITY_CAPABILITIES` structure. The desired capability is then translated to another, computer-specific SID in line 8. Since the AppContainer operates on ACLs, the SID of the capability is wrapped in a `SID_AND_ATTRIBUTES` structure along with the attribute

---

[15] https://github.com/dahall/Vanara

`SE_GROUP_ENABLED`, that allows the SID to be included in access checks.

The final step before the sandboxed process can be created is the initialization of a list of attributes that is added to the `STARTUPINFOEX` structure. Listing 4.4 shows the creation of a process attribute list, which is then updated with the security capabilities created in listing 4.3.

Listing 4.5 shows how the `STARTUPINFOEX` structure is then used to pass the capability information to the new process. It is created under the identity of the current user, albeit with a low integrity level.

To verify whether an application is indeed running in an AppContainer, tools like Process Explorer[16] can be used to check the group membership of a sandboxed process. In this example, the Notepad program shipped with Windows was started using the sample implementation of an AppContainer from this section. As shown in figure 4.4, Process Explorer reports the integrity as *AppContainer*, and the group *Mandatory Label* shows that the process runs on a low integrity level governed by MAC.

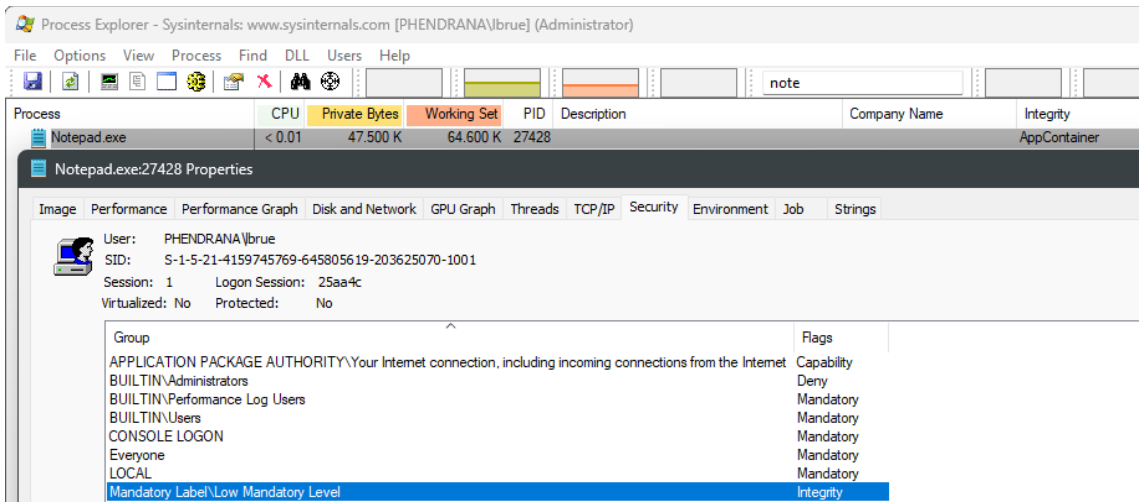

Figure 4.4: Process Explorer showing a sandboxed instance of Notepad

While the user can normally write text in this Notepad instance, trying to open a document or save the current text fails as expected. Since no explicit file or folder access was granted to the process, it is unable to even access the *Documents* directory of the current user, as shown in figure 4.5a. Trying to navigate to other directories or volumes fails with the same error message. By default, the only accessible folders are the working directory of the sandboxed application (with at least read access), and some common environment directories like `%temp%`, which are redirected to a folder specific for the container so that it can't interfere with non-sandboxed applications, as well as specific locations in the Windows Registry [90]. In comparison, a non-sandboxed instance of Notepad, shown in figure 4.5b, has regular access to the filesystem. Note that the

---

[16] An application that lists detailed properties of running processes — https://docs.microsoft.com/en-us/sysinternals/downloads/process-explorer

```csharp
1   private void SetSecurityCapabilities(ref Kernel32.SECURITY_CAPABILITIES capabilities,
    ↪   AdvApi32.SafeAllocatedSID appContainerSid, AdvApi32.WELL_KNOWN_SID_TYPE capabilitySid)
2   {
3       capabilities.AppContainerSid = appContainerSid.DangerousGetHandle();
4       var capabilitiesBuffer = new
    ↪   SafeHandleBuffer(Marshal.SizeOf(typeof(AdvApi32.SID_AND_ATTRIBUTES)));
5       var sidSize = (uint)AdvApi32.SECURITY_MAX_SID_SIZE;
6       var safePsid = new AdvApi32.SafePSID(sidSize);
7
8       AdvApi32.CreateWellKnownSid(capabilitySid, IntPtr.Zero, safePsid, ref sidSize);
9
10      var attributes = new AdvApi32.SID_AND_ATTRIBUTES
11      {
12          Attributes = (uint)AdvApi32.GroupAttributes.SE_GROUP_ENABLED,
13          Sid = safePsid.DangerousGetHandle()
14      };
15
16      Marshal.StructureToPtr(attributes, capabilityBuffer.DangerousGetHandle(), false);
17      capabilities.Capabilities = capabilitiesBuffer.DangerousGetHandle();
18  }
```

Listing 4.3: Building a capability from a SID

```csharp
1   private void SetProcessAttributes(ref Kernel32.STARTUPINFOEX startupinfo,
    ↪   Kernel32.SECURITY_CAPABILITIES capabilities)
2   {
3       var capabilitySize = Marshal.SizeOf(capabilities);
4       var capabilityBuffer = new SafeHandleBuffer(capabilitySize);
5       var processAttributeList = Kernel32.SafeProcThreadAttributeList.Create(
6           Kernel32.PROC_THREAD_ATTRIBUTE.PROC_THREAD_ATTRIBUTE_SECURITY_CAPABILITIES,
    ↪   capabilities);
7
8       Marshal.StructureToPtr(capabilities, capabilityBuffer.DangerousGetHandle(), false);
9
10      Kernel32.UpdateProcThreadAttribute(
11          processAttributeList.DangerousGetHandle(),
12          0,
13          Kernel32.PROC_THREAD_ATTRIBUTE.PROC_THREAD_ATTRIBUTE_SECURITY_CAPABILITIES,
14          capabilityBuffer.DangerousGetHandle(),
15          capabilityBuffer.BufferSize);
16
17      startupinfo.lpAttributeList = processAttributeList.DangerousGetHandle();
18  }
```

Listing 4.4: Adding the capabilities to the process attribute list

```
1  private void CreateSandboxedProcess(ref Kernel32.STARTUPINFOEX startupinfo, string processName)
2  {
3      using var currentIdentity = WindowsIdentity.GetCurrent();
4      using var currentToken = new GenericSafeHandle(currentIdentity.Token, Kernel32.CloseHandle,
           false);
5
6      var res = AdvApi32.CreateProcessAsUser(
7          new HTOKEN(currentToken.DangerousGetHandle()),
8          processName, null, null, null, false,
9          Kernel32.CREATE_PROCESS.EXTENDED_STARTUPINFO_PRESENT,
10         null, null, in startupinfo, out var processInfo);
11 }
```

Listing 4.5: Creating the sandboxed process

sandboxed version has so little access to the system, that it doesn't even have permission to retrieve the names of the volumes under *This PC*.

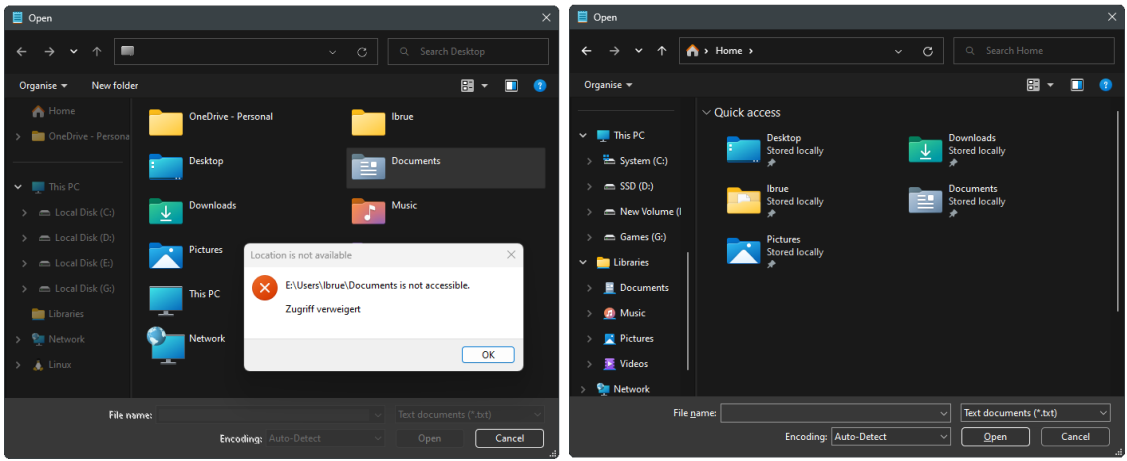

(a) Open dialogue in a sandboxed instance of Notepad

(b) Open dialogue in a regular instance of Notepad

Figure 4.5: *Open* dialogue of Notepad in a sandboxed and non-sandboxed instance

Unless the application should have intentionally only minimal access to the file system, it needs to be granted explicit permissions to read and write other files and folders. Listing 4.6 shows how the ACL of a file or folder can be modified to give full permissions to an AppContainer.

```csharp
private void GrantObjectAccess(AdvApi32.SafeAllocatedSID appContainerSid, string path)
{
    var type = AdvApi32.SE_OBJECT_TYPE.SE_FILE_OBJECT;
    var access = new AdvApi32.EXPLICIT_ACCESS
    {
        grfAccessMode = AdvApi32.ACCESS_MODE.SET_ACCESS,
        grfAccessPermissions = ACCESS_MASK.GENERIC_ALL,
        grfInheritance = AdvApi32.INHERIT_FLAGS.SUB_CONTAINERS_AND_OBJECTS_INHERIT,
        Trustee = new AdvApi32.TRUSTEE
        {
            MultipleTrusteeOperation = AdvApi32.MULTIPLE_TRUSTEE_OPERATION.NO_MULTIPLE_TRUSTEE,
            ptstrName = appContainerSid.DangerousGetHandle(),
            TrusteeForm = AdvApi32.TRUSTEE_FORM.TRUSTEE_IS_SID,
            TrusteeType = AdvApi32.TRUSTEE_TYPE.TRUSTEE_IS_WELL_KNOWN_GROUP
        }
    };

    AdvApi32.GetNamedSecurityInfo(path, type, SECURITY_INFORMATION.DACL_SECURITY_INFORMATION, out var ppsidOwner, out var ppsidGroup, out var ppDacl, out var ppSacl, out var ppSecurityDescriptor);
    AdvApi32.SetEntriesInAcl(1, new[] { access }, ppDacl.DangerousGetHandle(), out var newAcl);
    AdvApi32.SetNamedSecurityInfo(path, type, SECURITY_INFORMATION.DACL_SECURITY_INFORMATION, ppsidOwner, ppsidGroup, newAcl.DangerousGetHandle(), ppSacl);
}
```

Listing 4.6: Adding the SID of the container to the ACL of a file or folder

**Evaluation**

AppContainers enable any application to be run as a process with low integrity level. Since regular users operate by default on a medium integrity level, and Windows implements mandatory access control, low IL processes are effectively prohibited by interfering with objects with a higher IL. In comparison to CAS and AppDomains, AppContainers are not limited to managed code and are subject to operating system-level security enforcement by being integrated into existing ACLs of objects. Their integration as part of widely used software like Chromium gives credibility to their security in real-world applications.

The disadvantage comes with their complicated creation for applications that are not automatically containerized (like Windows Store apps are). The lack of a managed API requires use of sparsely documented and difficult to implement native Windows functions. Access to files and folders that are required for the normal operation of a sandboxed application has to be assigned manually. Granting access to entire volumes[17] is more difficult, as their ownership usually lies with the operating system, and requires elevated permissions to alter them. While existing applications can be put into a container, it may impede their functionality as they assume to be run at medium integrity level.

Applications that are not designed to run at low IL, especially if their source code is unavailable or not readily modifiable, have to be analysed for possible locations that they are allowed to access. Communication with processes of higher IL is, intentionally due to the usage of MAC, another issue, as a low IL process cannot simply call more trusted processes and has to communicate via other means, such as IPC techniques. Microsoft generally recommends distrusting low IL processes [90]. If a sandboxed application deals with critical or sensitive information, or is prone to attacks from malicious actors, incoming and outgoing data has to be carefully validated to prevent circumvention of the container.

For example, a low IL process could create an object in shared memory (such as a named pipe, or a memory-mapped file) that it knows will be opened by a process with higher IL. The low IL process could fill the object with data that will cause, if not properly validated, the higher IL process to crash or execute arbitrary code, thus breaking out of the limitations of its integrity level. This type of attack, also called a *squatting attack* [133], requires knowledge about the vulnerable process and also shows that, depending on the threat model, multiple levels of security may be necessary.

## 4.6 Summary and discussion

Chapter 4 provided a general overview of process isolation and approaches on how applications can run alongside each other while minimizing accidental or malicious interference. In the context of LCDPs, a threat model was introduced that served as the basis for the evaluation of two concrete implementations of OS-level virtualization. Additionally, three currently available mechanisms for code security in C# were discussed, where AppContainers, which use Windows' MIC, were the most viable and widely used option.

---

[17] For example, a partition such as `C:`

This chapter served as a basis for answering RQ III, and, in a broader sense, also RQ II. While supporting CDs may help to reduce usage of potentially insecure code, there is always the possibility that certain APIs are missed or advice is ignored. Additional security on a lower, OS-level, therefore also alleviates security problems that slipped through preventive measures on a higher, user-code level. Implementing OS-level security requires additional infrastructure and systems to deploy and run the applications to be secured. Section 4.4.2 detailed the usage of Docker for the deployment of a small ASP.NET Core application, which required at minimum a virtualization-capable server and the effort to create a Dockerfile to build the application and install all their dependencies. At scale, additional services to orchestrate the instantiation and management of containers are needed.

While there are different APIs to provide code security in .NET are available, the only currently viable and supported implementation are AppContainers, which limit the access of applications to the rest of the system using integrity levels. AppContainers are implemented in widely used applications such as Chromium, but are difficult to implement and configure correctly. They also require bootstrapping, that is, the sandboxed process has to be specifically started inside an AppContainer, which has to be done by a separate process. Configuration involves setting the desired security capabilities, which are not well documented, and controlling access to files and folders via an ACL. While any application can be placed in an AppContainer, imposing external limits can lead to unexpected behaviour if the application is not aware that access to normally available resources is suddenly restricted.

## Chapter 5

# Maintenance and management of custom code

Mechanisms for the support of citizen developers when *writing* code were presented in chapter 3, and implementations for securely *running* that code were discussed in chapter 4. This chapter introduces two topics that are routinely part of the maintenance and management of *existing* code: debugging and versioning.

## 5.1 Debugging

Debugging is an indispensable activity in software development that allows developers to inspect a particular piece of code and analyse the state of the program during runtime [160]. Modern IDEs such as Visual Studio or IntelliJ IDEA include extensive support for debuggers that allow the user, for example, to place *breakpoints* at specific lines in the code to pause execution and examine the contents of variables and step through the code statement by statement. However, a much simpler and widely adopted technique colloquially called *printf debugging* is still often employed as the first step in debugging code issues [11]. Printf debugging, named after the set of `printf()` functions in the C programming language, involves inserting statements that output information during program runtime, such as the values of variables or simple messages indicating the control flow. The developer then observes the output and can reason about the state of the program at the time of the output. The advantage of this technique is that it is virtually always available, as most programming languages support some form of output, and requires no additional tools or familiarization with a more complex debugger.

### 5.1.1 Usage in low-code applications

Debugging low-code applications is considered as a challenge by developers [3, 75], as the focus on graphical development environments makes it more difficult to employ traditional means of debugging. Generated applications can still produce code that can be debugged with external tools, for example with Eclipse[1] in the case of Mendix's *Java Actions* [80]. However, this requires development experience and would likely involve additional training for citizen developers.

Another available piece of information for debugging is the output of the compiler that builds an application. In the context of generated LCAs, this output is available on at least two different occasions: when the entire LCA is compiled, and when a user writes additional code for the application. The latter case requires that the LCDP supports selective compilation of user-written code. The compiler output of the entire generated LCA is likely not helpful to the CD, since it will contain messages about all parts of the application, even those the CD never worked on, and probably can't fix – for example because it's a bug in the code generator. Compiler output is also generally fairly technical, and geared towards experts [137].

Limiting the compiler output to the code the CD has written and currently debugging has, independently of the difficulty of understanding it, some advantages. For one, it provides more focused messages, as compared to building a whole application, a smaller amount of code is compiled. It also enables a much faster feedback loop. Users can apply changes to the code more frequently and can experiment more, as they don't have to wait until the application is recompiled, since compiling only several lines of code is faster than rebuilding the entire LCA.

Enhancing compiler output messages seems like an obvious way to improve their comprehensibility for citizen developers. The effectiveness of *enhanced compiler error messages* (ECEMs) is, while studied, not conclusively answered. Multiple studies [33, 119, 127] suggest that ECEMs have little effect in aiding novices when encountering errors. However, they contributed in reducing the most commonly received Java compiler errors for a group of students [10], and were generally found helpful by participants [129].

As previously mentioned, *printf debugging* is still frequently used by experienced developers and easy to learn for novices. Providing citizen developers with the facilities to show the results of simple output statements gives them the opportunity to debug their code with minimal teaching effort. Additional written code for SCOPELAND is usually limited to small functions and statements applied to the current page, without complex object hierarchies [120, p. 47]. Thus, the absence of more advanced debugging options is compensated by smaller and simpler amounts of code. Section 5.3.1 details the implementation and usage of compilation and debugging features for small C# scripts in an application.

---

[1] A Java-based IDE for a wide variety of programming languages — `https://eclipseide.org`

## 5.2 Versioning

Another integral part of software development is the usage of version control systems (VCSs) [58], such as Git, Subversion or Perforce. VCSs allow developers to store and manage changes to a code base. Changes to files and folders are incrementally *committed* or *checked in*, along with metadata such as the author's name, a timestamp, and a message, into either a central or distributed *repository*. Maintaining a VCS allows developers to view a history of changes and switch between different revisions as necessary, for example to compare modifications to a file or go back to a previous version. Additionally, VCSs enable collaboration between multiple developers or teams across a single code base. New features or bugfixes can be worked on independently, without affecting others, until the implementation is finished and can be *merged* into the central repository. Conflicting changes can be detected, revised and resolved before potentially inconsistent code is committed to the shared code base.

### 5.2.1 Usage in low-code applications

Studies indicate that VCSs are difficult to use, even for professional developers. Yang et al. [159] evaluated the difficulties associated with Git, and find that trained developers still can have issues using Git and that further research in assisting them is needed. Isomöttönen et al. [63] describe the experience of introducing Git to computer science students and conclude that the challenges encountered are independent of the concrete VCS. They report that most of the problems the students ran into were related to the intricacies of the Git command-line syntax, and difficulties in seeing the advantages of a rather complicated VCS in short-term coursework usage with a lack of authentic use cases. As both skilled developers and computer science students can struggle in employing a VCS as part of their work, citizen developers will require assistance to version their applications.

Version control systems usually provide a command line interface to interact with them. Applications with GUI support are available as well, often from third parties. Git, for example, is directly shipped with git-gui[2]. External tools can provide integration with other services, such GitHub Desktop[3] for GitHub or GitKraken[4] for a variety of issue-trackers. Integration in development tools and IDEs is often specifically shipped as part of them, or is available through add-ons. Unless a LCDP directly supports integration of a VCS (such as Mendix [83]), or provides an API to do so (like OutSystems [124]), a citizen developer would have to manually manage versioning their applications. Besides the inherent difficulty of using a VCS that was described previously, this might not even be possible, as the end-user has no access to the inner structure of a LCDP. While the source code may be available as part of generated LCAs (see section 2.2.3), this may not be the case in hosted applications. As such, if a LCDP supports a VCS, its usage needs to be abstracted for a citizen developer.

---

[2] https://git-scm.com/docs/git-gui
[3] https://desktop.github.com
[4] https://www.gitkraken.com

Another point to consider is *what* to version. The application components developed in LCDPs can be stored by the platform itself. For example, the pages developed in SCOPELAND are not stored as code, but as structured metadata inside the internal meta database [120, p. 48]. Any additionally written code is also embedded in this metadata [120, p. 47]. In this case, the application components (pages) exist as an entity in a database and can be managed and versioned by the LCDP. For instance, it can offer the CD to open a previously saved version of the application. The issue with this approach is, that the code is not stored independently of the rest of the LCA. To look at a previous version of their code, users have to go back to a previous version of the page. Changes to the other content of the page are thus also reverted. As design aspects of a page can be stored globally with reusable stylesheets in SCOPELAND [120, p. 55], several visual elements, such as colours and font settings, are already independent of the page. A similar pattern can be applied to the code embedded in the pages. Integrating a simple VCS into the LCDP enables citizen developers to gain quickly access to previous versions of their code, along with a history of changes, without having to go back to a previous version of the entire page. An example application showing the implementation and usage of such a system is shown in section 5.3.2.

## 5.3 Implementation

To demonstrate the feasibility of providing citizen developers with basic debugging facilities and a simplified version control system, the author implemented a *Code Management Sample* application (CMSA) that is open source and publicly available on GitHub [18].

The application consists of two major features. The first is the ability to compile and execute console-based C# code, without requiring any external tools, and giving the CD access to the compiler- and console output for debugging purposes. The other allows storing and retrieving the currently active code in a Git repository without the need to learn any specific Git commands. The history of the file is displayed in a simple table, and older versions can be previewed and restored as necessary.

### 5.3.1 Debugging

Running and debugging the user-written scripts requires compilation of the code. As the citizen developers cannot be expected to develop their code in a separate environment (like an external IDE or editor), the compilation needs to happen as part of their usual workflow. On-demand compilation of code during the runtime of a program can be achieved with the APIs provided by the Roslyn compiler, specifically with its scripting functions of the

`CSharpScript` class [99]. A simplified example is shown in listing 5.1.

```
public async Task CompileAndRunScript()
{
    string sourceCode = "System.Console.WriteLine(\"Hello World\");";
    var script = CSharpScript.Create(
        sourceCode,
        ScriptOptions.Default
            .WithReferences(Assembly.GetAssembly(typeof(Console))));
    script.Compile();
    await script.RunAsync();
}
```

Listing 5.1: Compiling and executing a script with Roslyn

Line 3 contains the code that is compiled and executed, in this case `Hello World` is printed to the console output. In the following lines, a `Script` instance is created, and the required references, assemblies that are used in `sourceCode` (the `System.Console.dll` assembly from .NET in this example), are added. The script is compiled in line 8 and asynchronously executed in line 9.

The `System.Console.WriteLine()` method call used in this listing can be used for printf debugging, as it writes the passed arguments to the console. Running this example doesn't produce any visible output, as no console window is open. To capture the console's contents, the *standard output* has to be redirected to an object that can be used by the application to provide it to the user. Listing 5.2 shows how the standard output can be captured by redirecting it to a `StringWriter`, a .NET class to sequentially write data to a string.

```
try
{
    await using var writer = new StringWriter();
    Console.SetOut(writer);
    ScriptState<object> scriptResult = await Script.RunAsync();
    await writer.FlushAsync();
    var consoleOutput = writer.GetStringBuilder().ToString();
}
finally
{
    Console.SetOut(Console.OpenStandardOutput());
}
```

Listing 5.2: Capturing and redirecting console output to a string

In line 4, the console output is redirected to an instantiated `StringWriter`. Afterwards, the script is executed, and following its completion, `writer.FlushAsync()` forces remaining buffered data to be written to the `StringWriter`. Any text that is written during the script execution is now present in the `consoleOutput` string. Finally, the default standard output is restored.

This captures the console output, but there are two more sources of information that need to be considered: exceptions (errors) and return values. If an unhandled exception occurs during debugging of an .NET application, the execution is halted and the developer is notified of the error. To check if an error occurs, the `ScriptState` object, returned upon execution of `Script.RunAsync()`, provides an `Exception` property that contains information about the error. This needs to be passed to the user as well. Otherwise, the script will simply exit without notifying the user. Similarly, the return value, provided by any top-level `return` statements in the script, can be retrieved from the `ReturnValue` property of the `ScriptState` object.

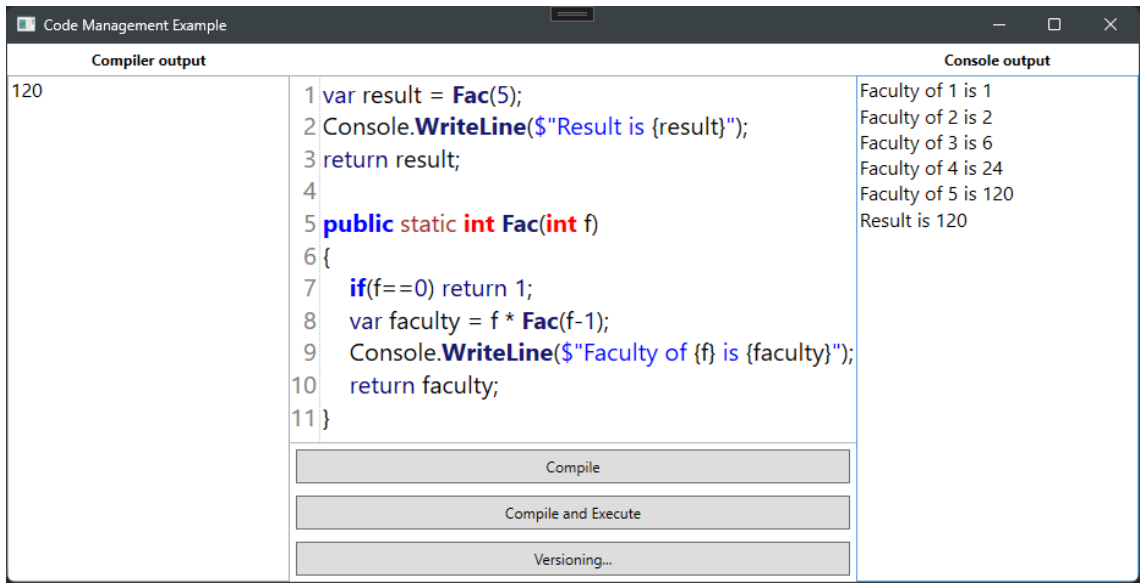

Figure 5.1: Debugging example showing console and return value output

Figure 5.1 shows the example application implemented for this chapter. In the centre column, the citizen developer can input C# code. This example contains a simple script to calculate the faculty of a number. Upon clicking the middle (*Compile and Execute*) button, the script is built and run. Any strings emitted by `Console.WriteLine()` statements are printed to the text area in the right column. The return value, provided by the statement in line 3 is printed to the text area in the left column.

In case of an error in the script, the output provided by the compiler is also printed in the left column. In figure 5.2, a spelling error (`retrn` instead of `return` in line 3) was made in the code, and the resulting compiler error message is displayed on the left.

The interface of the example application is deliberately designed to be as minimal and easily understandable as possible. As previously mentioned in section 5.1.1, the scripts written for LCA are expected to be relatively small. Compilation is therefore expected to be quick as well, and the user is provided with both a button to only compile the code, and check it for errors, and to compile and immediately run it.

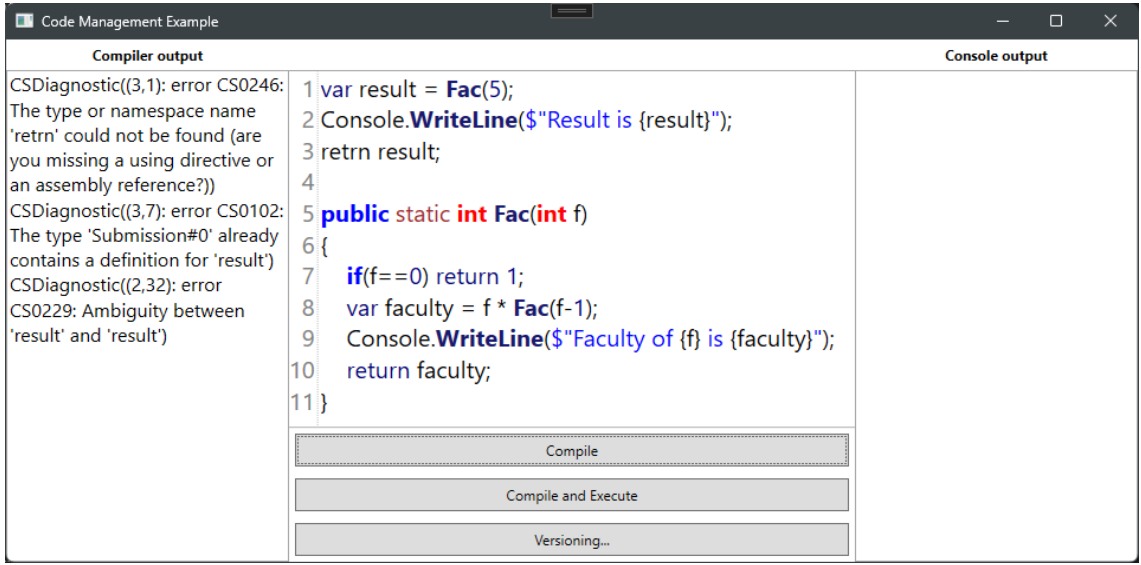

Figure 5.2: Debugging example showing compiler error output

However, citizen developers are unlikely to exclusively write trivial scripts like the faculty calculation algorithm shown in figures 5.1 and 5.2. The code that is added to low code applications covers special cases that can't be expressed by the LCDP by default. These can be, for example, integrations with other applications, connectivity with external services [120, p. 69] or special parsing logic for data present in the LCA. As such, being able to immediately execute scripts might not be beneficial in all cases and can depend on the context. Suppose the script depends on external data, for example from a web service or a database, that is passed during runtime of the application. While the citizen developer can run and test the script isolated in the editor, it can only be fully integrated when it is run as part of the entire LCA.

Another aspect to consider are the external libraries that are available to the CD. In listing 5.1, only the `System.Console` assembly is made available to the editor. In a more sophisticated application, many more assemblies may be required, such as for database access or the wrapped APIs introduced in section 3.4.1. While more references can be added to Roslyn either by default or at the runtime of the LCA, this needs to be done as transparently as possible for the CDs, as deciding what additional references they need is not within their usual expertise.

### 5.3.2 Versioning

As explained in section 5.2.1, versioning control systems can be difficult to use. A simple abstraction layer, that internally operates on a Git repository, was developed as part of the CMSA and is shown in figure 5.3. This provides CDs with basic versioning features without having to learn the intricacies of a VCS.

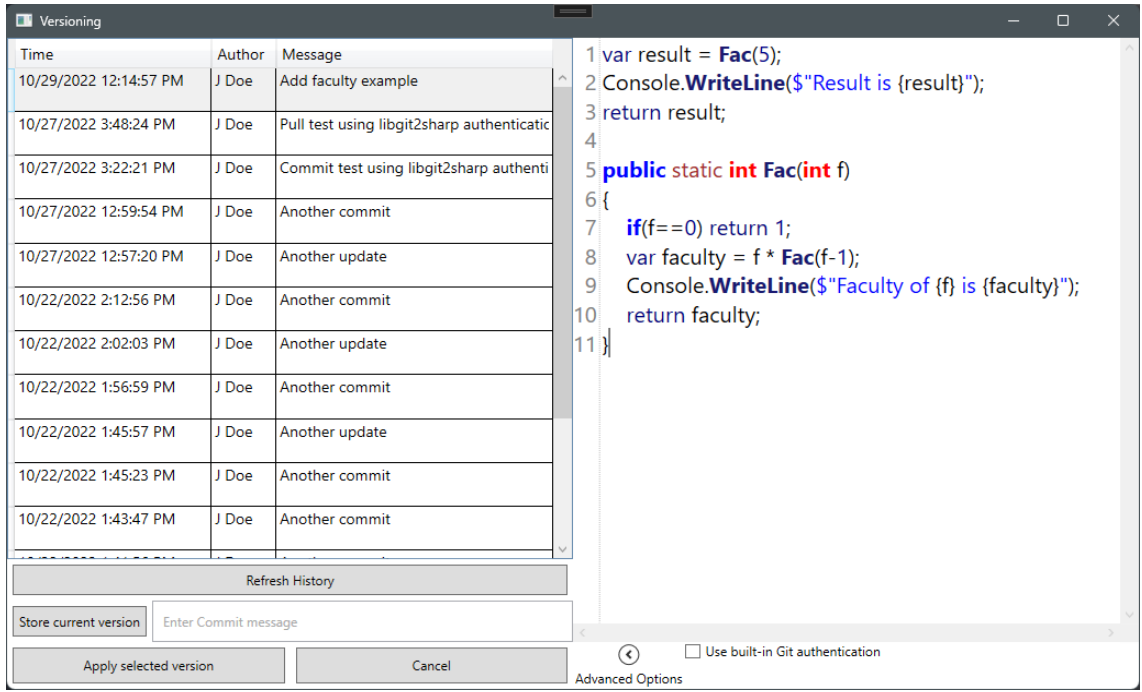

Figure 5.3: Versioning window of the CMSA

The example consists of three parts: A history browser on the left side shows all previous versions of the script, along with a timestamp, the author, and a commit message. Below are several controls to store and restore versions to and from the repository. On the right side is a preview area that shows the content of the script from the selected version in the history browser. The user can browse through the different versions via the history, and preview and restore a specific version. They can also store the current version of the script, and enter a message that details the changes and will appear as part of the history.

The abstraction layer is implemented with *libgit2sharp*[5], a library enabling access to Git repositories from .NET, and several direct invocations of the Git executable, to work around authentication limitations in libgit2sharp.

The CMSA operates on a single, existing Git repository. There are no special requirements towards the repository, it can be either used locally, or connected to a remote host with internal or external services such as GitHub or Bitbucket[6]. As the script does not directly exist as a file on the local filesystem, it needs to be packaged in a format that can be stored in the repository. As Git stores its data internally in a content-addressable filesystem, essentially a key-value store [23], the script doesn't *need* to exist as a file. The CMSA serializes the script into a JSON object, using .NET's `JsonSerializer` class, and stores it as a *blob* [23] in the repository. Each script is assigned a GUID that is used to identify and address its blob representation in Git's filesystem.

---

[5] https://github.com/libgit2/libgit2sharp

[6] A hosting service for source code, similar to GitHub - https://bitbucket.org

```
public record VersioningModel(string Author, string Email, Guid BlobId, string Script, string
    RepositoryPath);

public void CommitCode(VersioningModel model)
{
    var blob = SerializeBlob(model.BlobContent);

    using var repo = new Repository(model.RepositoryPath);

    var committer = new Signature(model.Author, model.Email, DateTime.Now);

    repo.Index.Add(blob, model.BlobId.ToString(), Mode.NonExecutableFile);
    repo.Index.Write();
    repo.Commit(CommitMessage, committer, committer);
}
```

Listing 5.3: Committing code using libgit2sharp. Based on [70]

Listing 5.3 shows how a blob is created and committed to the repository. The parameter `model` is a `record`, its definition shown in line 1, containing context-specific information about the repository, such as the path to the repository or author of the commit. After the blob is serialized and the repository initialized, a `Signature` object is instantiated to connect the identity of the author of the commit with the commit itself. Lines 11 and 12 add the blob to Git's index, also called the *staging area*, where changes are collected, before line 13 commits them to the repository.

At this point, the blob is only present in the local repository. Since Git is a distributed VCS, an additional remote repository can be configured. This allows multiple developers to *push* their work to a single, centralized point on a remote server, which is essential for collaboration. A public example repository was set up for the CMSA by the author and can be found on GitHub [19]. When multiple developers work with a common repository, some kind of user management and authentication is necessary [24]. Thus, operations on the remote repository, such as pushing changes to and pull changes from it, requires the user to authenticate with the server the repository resides on. Integration in existing enterprise authentication is out of scope for this thesis, however libgit2sharp generally also supports simple authentication with username and password. As hardcoding and exposing these in the source code presents a significant security issue, the CMSA relies on the *Git Credential Manager* for Windows [25], that can be installed as part of a standard Git installation and uses Windows' built-in credential management. After the credentials for a certain server are entered once and stored, Git can use them implicitly when communicating with a remote repository.

As there is no native API in .NET to access the credential manager, the library CredentialManager [152] is used in the CMSA, which provides a wrapper of the native API [112]. The credentials are identified by an application-dependent key. The Git Credential Manager uses a format of {namespace}:{service} [49], where `namespace` defaults to `git` and service consists of the URL of the host of the repository. A full key might look like, for example, `git:https://github.com`.

```
1   private static NetworkCredential GetOrAskCredentials(string credentialIdentifier)
2   {
3       var existingCredentials = CredentialManager.GetCredentials(credentialIdentifier);
4
5       if (existingCredentials ≠ null) return existingCredentials;
6
7       var rememberCredentials = false;
8       return CredentialManager.PromptForCredentials(credentialIdentifier, ref rememberCredentials,
    ↪    $"Enter credentials to access {credentialIdentifier}.", "Credentials Request");
9   }
```

Listing 5.4: Retrieving credentials from the Windows credential store

Retrieving the credentials is shown in listing 5.4. If the credentials with the specified key are not already present in the credential store, the user is asked to provide them. The stored username and password can then be used with libgit2sharp to authenticate requests to the remote repository, which is shown in listing 5.5

It should be noted that the `NetworkCredential.SecurePassword` property used for the password in listings 5.4 and 5.5 is of the type `SecureString`, which tries to minimize the exposure of sensitive data during runtime of an application. While it is more secure than a regular variable of type `string`, there are still instances where the password may be exposed in memory [100]. If the risk of exposing passwords in memory is deemed too large, authentication needs to be conducted outside the application, since there is no alternative to `SecureString` in .NET. Another option may be to not use libgit2sharp for communication with the remote repositories, and instead work with Git and its configured authentication directly, by invoking its executable with the appropriate arguments (such as `git pull` or `git push`). This is implemented in the CMSA as well and can be enabled via the advanced options shown in figure 5.3.

```
 1  private static void Push(NetworkCredential credentials, Repository repo)
 2  {
 3      var options = new PushOptions
 4      {
 5          CredentialsProvider = (_, _, _) =>
 6              new SecureUsernamePasswordCredentials
 7              {
 8                  Username = credentials.UserName,
 9                  Password = credentials.SecurePassword
10              }
11      };
12      repo.Network.Push(repo.Head, options);
13  }
```

Listing 5.5: Pushing changes to a remote repository with libgit2sharp

## 5.4 Summary and discussion

Chapter 5 demonstrated how techniques that are common in traditional software development processes can be integrated in LCDPs. These techniques require experience and familiarization with tools and can be challenging to learn even for professional developers. The CMSA shows how debugging and versioning tools can be used in a simplified way to enable citizen developers to benefit from them without the need for extensive training.

The results from this chapter contribute towards an answer to RQ I, namely that assistance in debugging and versioning, by providing limited and clearly defined access to these tools, can be a viable way of supporting CDs. The CMSA only presents some of the ideas that could be implemented in a LCDP. The concrete implementation depends on external factors, namely the requirements of a LCDP towards debugging and versioning, the amount of technical details the target group of citizen developer can be exposed to, and the underlying technology stack.

**Debugging**

While printf debugging is a simple and widely used method to analyse issues with code, it is also the only option the CMSA offers. Other debugging techniques, such as breakpoints, inspecting the call stack or watching local variables are not supported. Of the previously discussed LCDPs that offer code extensions, both OutSystems [43] and Mendix [79] only support debugging via an external IDE. Microsoft's Power Apps platform provides the *Monitor* tool [89], that enables detailed tracing of events that occur in the LCA, which is essentially a more advanced and automated version of debugging with output statements.

Depending on the used libraries in the script, the assemblies containing them need to be added as references to Roslyn's scripting engine. Although this is easily done from code (shown in listing 5.1), this means that either all references have to be hard-coded before use (potentially adding too many or too few libraries), or added on demand during runtime (which requires someone with that specific knowledge, or shifts the responsibility to the citizen developers).

Compiler errors are returned and shown to the CD verbatim, without any special processing. As shown in section 5.1.1, it is inconclusive whether enhanced compiler output messages are helpful to users or not. A compromise may be to add more helpful messages incrementally for the most encountered compiler errors, based on real-world experience. The augmentation framework introduced in section 3.3 could be used to visualize errors directly in the editor.

**Versioning**

The CMSA is currently limited to a single script that can be edited and versioned. An actual application would have many scripts that need to be managed individually. The blob representing the code is stored with a unique, but otherwise nondescript ID in the repository, making usage outside the CMSA inefficient. Since the repository is not specific to the application, it can be used from external Git clients and services as well. This necessitates a better storage system with more descriptive names and a file and folder structure that is both human-readable for access from outside clients and can be reliably interacted with from external services. For example, the repository could be integrated into an existing issue-tracking system, connecting commits with task and bug tickets, or keeping a reference to the blob from a database, to map a script to a certain element of the LCA.

The Git repository also has to be set up beforehand, including initialization and potentially hosting on a remote server; the CMSA cannot initialize a new or clone an existing repository. Git also has to be installed on all client machines or the server (depending on the type of LCDP) and prepared for multiple users. Authorization must be configured as well, so that commits are distinguishable between users.

Collaboration between multiple developers inadvertently results in merge conflicts, caused by overlapping edits to code [20]. These conflicts have to be resolved before code is pushed or pulled to or from the repository. When Git encounters a change that cannot be automatically resolved, it will pause the merge operation and wait for the user to choose which of the conflicting lines of code are to be used. The CMSA doesn't offer any special facilities to handle merge conflicts, they have to be resolved via other means, such as the command line or an external tool. Since the CMSA is primarily used for demonstration purposes, conflicts are not likely unless changes are made to the repository from the outside. A LCDP would, however, require a user-friendly option to show and resolve conflicts that can be handled by citizen developers.

## Chapter 6

# Conclusions and future work

## 6.1 Summary

This thesis presented approaches to extend code editors in LCDPs to support their users with context-based visual augmentations. These augmentations can be used, for example, to detect potentially insecure method calls and suggest more secure alternatives, or highlight statements based on regular expressions, to provide the user with additional feedback or information. To securely run different LCAs concurrently on a server, currently available technologies to isolate applications from each other on OS- and process-level were evaluated.

The results developed during this thesis can be successfully used to answer the research questions posed in section 1.4:

> *RQ I: How can citizen developers in low-code environments that offer C#-based extension of their applications be supported?*

The augmentation framework created as part of chapter 3 provides customizable visual clues on a syntactic level. Augmentations can change the appearance of predefined statements, based on strings or regular expressions, by altering their typographic properties, displaying geometric elements or images, or replacing them with more meaningful identifiers. The code itself remains unchanged, the display is purely visual. On a semantic level, an approach of integrating custom Roslyn analyzers and code fixes into the RoslynPad code editor control, embeddable into any WPF application, was demonstrated. Chapter 5 provided examples how debugging and version control, two commonly used techniques in traditional software development, can be simplified and integrated in LCDPs in ways that are easy to use and understand for CDs.

*RQ II: How to guard against malicious or unintended harmful C# code that is written to extend low-code applications?*

Detecting potentially insecure or harmful code can be achieved by detecting specific statements, types or, method calls with either the augmentation framework or Roslyn analyzers, and emitting warnings or hints when found. However, the targeted APIs have to be known in advance and need to be routinely administered and updated. Wrapping complicated or security-critical libraries with a custom API can reduce the chance for misconfiguration of services. Neither of these will protect against actors that try to inject malicious code into a LCA. The examined API-level protections, CAS and AppDomains, are considered obsolete and offered no full protection in the first place. The threat model used in section [4.2](#) assumes that an attacker tries to compromise other applications hosted on the same server, and since there are no viable code-level security solutions, the alternative is to isolate applications against each other and the system on a lower level, which leads to RQ III:

*RQ III: How can the execution of low-code applications be isolated in a way, that additionally written C# code cannot adversely affect foreign resources on a system?*

For this purpose, both AppContainers and OS-level virtualization with Docker offered adequate options, the choice depends on the kind of applications (with or without UI) that need to be run, as well as requirements for scalability and platform-independence. AppContainers need to be bootstrapped by a separate process and are limited to Windows operating systems and applications. Since they can be configured to prevent sandboxed applications from using defined system services or only allowing access to certain files and folders, this is best suited for applications where these usages are known or can be altered to respond accordingly. Otherwise, applications may exhibit undesirable behaviour, when they don't assume that they are run with limited integrity and encounter unexpected restrictions when accessing the system. Docker can be used to provide applications with a customizable, deterministic and containerized environment, while also limiting their access to the underlying operating systems and its resources. This enables multiple containers to be run on the same operating system concurrently, without the need for a completely virtualized operating system in a VM. Containers can be integrated into orchestration engines that provide management and scalability features for multiple containers. While Docker can work on multiple operating systems, it is primarily suited to run server-side applications without a GUI.

## 6.2 Outlook

Since the augmentation framework currently targets only a single editor component, a natural extension is the adaptability to other code editor controls. As the extensibility points of other editors can be expected to be different, more limited or less restricted, the concrete implementation for an editor could be abstracted away from an augmentation. It would be reduced to a simple, data-holding class containing only the information describing how an augmentation should look to a user. The concrete editor implementations for rendering the augmentations reside in encapsulated classes, and are independent of the builder methods to configure an augmentation. The framework can also be expanded to support different kinds of input data, such as multiple image formats or importing analytics from an external sources, like databases.

Similarly, the CMSA from section 5.3 could be adapted to support other version control systems than Git. As long as the functionality is kept to a common feature set (history, pushing and retrieving data from the VCS), the difficulty lies primarily in the availability of native libraries of other VCS. An alternative can be the implementation of a wrapper for the command-line interfaces most VCS provide. Using different compiler platforms for languages apart from those supported by Roslyn (C# and Visual Basic .NET) is dependent on the targeted language and whether suitable tools are available. For example, interoperability with Python code is possible with IronPython [62].

# List of Figures

# List of Listings

# List of Tables